\documentclass[fleqn,twocolumn,times,baselineskip]{aastex63}

\usepackage{mathtools}

\renewcommand{\thesection}{\arabic{section}}
\usepackage{float}
\usepackage{subfig}
\usepackage{listings}
\usepackage{makeidx}
\usepackage{xcolor}
\usepackage[T1]{fontenc}
\usepackage{braket}
\usepackage{comment}

\newcommand\HI{H{\small{I}}}
\newcommand\HII{H{\small{II}}}
\newcommand\Lyaf{Ly$\alpha$ forest}

\accepted{February 1, 2022}
\shorttitle{Lyman-$\alpha$ forest lss mapping in real and redshift space}
\shortauthors{Sinigaglia et al.}
\graphicspath{{./}{figures/}}

\begin{document}


\title{Mapping Lyman-alpha forest three-dimensional large scale structure in real and redshift space}

\correspondingauthor{Francesco Sinigaglia}
\email{francesco.sinigaglia@phd.unipd.it}
\correspondingauthor{Francisco-Shu Kitaura}
\email{fkitaura@iac.es}

\author[0000-0002-0639-8043]{Francesco Sinigaglia}
\affiliation{Department of Physics and Astronomy, Università degli Studi di Padova, Vicolo dell’Osservatorio 3, I-35122, Padova, Italy}
\affiliation{INAF - Osservatorio Astronomico di Padova, Vicolo dell’Osservatorio 5, I-35122, Padova, Italy}
\affiliation{Instituto de Astrof\'isica de Canarias, s/n, E-38205, La  Laguna, Tenerife, Spain}
\affiliation{Departamento  de  Astrof\'isica, Universidad de La Laguna,  E-38206, La Laguna, Tenerife, Spain}

\author[0000-0002-9994-759X]{Francisco-Shu Kitaura}
\affiliation{Instituto de Astrof\'isica de Canarias, s/n, E-38205, La  Laguna, Tenerife, Spain}
\affiliation{Departamento  de  Astrof\'isica, Universidad de La Laguna,  E-38206, La Laguna, Tenerife, Spain}

\author[0000-0001-5028-3035]{Andr\'es Balaguera-Antol\'inez}
\affiliation{Instituto de Astrof\'isica de Canarias, s/n, E-38205, La  Laguna, Tenerife, Spain}
\affiliation{Departamento  de  Astrof\'isica, Universidad de La Laguna,  E-38206, La Laguna, Tenerife, Spain}

\author{Ikkoh Shimizu}
\affil{Shikoku Gakuin University, 3-2-1 Bunkyocho, Zentsuji, Kagawa, 765-8505, Japan}

\author[0000-0001-7457-8487]{Kentaro Nagamine}
\affiliation{Theoretical Astrophysics, Department of Earth and Space Science, Graduate School of Science, Osaka University, \\
1-1 Machikaneyama, Toyonaka, Osaka 560-0043, Japan}
\affiliation{Kavli-IPMU (WPI), University of Tokyo, 5-1-5 Kashiwanoha, Kashiwa, Chiba, 277-8583, Japan}
\affiliation{Department of Physics \& Astronomy, University of Nevada, Las Vegas, 4505 S. Maryland Pkwy, Las Vegas, NV 89154-4002, USA}

\author{Manuel S\'anchez-Benavente}
\affiliation{Departamento  de  Astrof\'isica, Universidad de La Laguna,  E-38206, La Laguna, Tenerife, Spain}

\author[0000-0002-5934-9018]{Metin Ata}
\affiliation{Kavli-IPMU (WPI), University of Tokyo, 5-1-5 Kashiwanoha, Kashiwa, Chiba, 277-8583, Japan}



\begin{abstract}
This work presents a new physically-motivated supervised machine learning method, \textsc{hydro-bam}, to reproduce the three-dimensional Lyman-$\alpha$ forest field in real and in redshift space learning from a reference hydrodynamic simulation, thereby saving about 7 orders of magnitude in computing time. We show that our method is accurate up to $k\sim1\,h\,\rm{Mpc}^{-1}$ in the one- (probability distribution function), two- (power-spectra) and three-point (bi-spectra) statistics of the reconstructed fields. When compared to the reference simulation including redshift space distortions, our method achieves deviations of $\lesssim2\%$  up to $k=0.6\,h\,\rm{Mpc}^{-1}$ in the monopole, $\lesssim5\%$ up to $k=0.9\,h\,\rm{Mpc}^{-1}$ in the quadrupole. 
The bi-spectrum is  well reproduced for triangle configurations with sides up to  $k=0.8\,h\,\rm{Mpc}^{-1}$. In contrast, the commonly-adopted \textit{Fluctuating Gunn-Peterson approximation} shows significant deviations already neglecting peculiar motions at configurations with sides of $k=0.2-0.4\,h\,\rm{Mpc}^{-1}$ in the bi-spectrum, being also significantly less  accurate in the power-spectrum (within 5\% up to $k=0.7\,h\,\rm{Mpc}^{-1}$). We conclude that an accurate analysis of the Lyman-$\alpha$ forest requires considering the complex baryonic  thermodynamical large-scale structure relations. Our hierarchical domain specific machine learning method can efficiently exploit this and is ready to generate accurate Lyman-$\alpha$ forest mock catalogues covering large volumes required by surveys such as DESI and WEAVE.

\end{abstract}



\section{Introduction} \label{sec:intro}

One of the most outstanding discoveries in modern Cosmology has been the acceleration of the expansion of the Universe \citep{Riess1998,Perlmutter1999}. This finding supported the so-called $\Lambda$ cold dark matter ($\Lambda$CDM) model and revived the existence of an accelerating force as opposed to gravity, called \textit{dark energy} \citep[see e.g.][and references therein]{Peebles2003}, initially postulated by Einstein's General Relativity. Since then, $\Lambda$CDM and other alternative theories have been put under test to unveil the nature of the accelerating expansion and probe the geometry of our Universe. To this end, the Large Scale Structure of the Universe \citep[LSS hereafter, see, e.g.][]{Springel2006} offers invaluable insights into these profound questions, carrying an imprint of the underlying fundamental physics which determines the evolution of the Universe and ultimately its fate.

While most of the efforts have been focused on mapping the Universe on large scales with galaxies at $z\lesssim2$, other complementary probes open the possibility to trace the LSS at higher redshifts.

In particular, the Lyman-$\alpha$ (Ly$\alpha$ hereafter) absorption features (dubbed \textit{Ly$\alpha$ forest}), originated in quasar spectra by interactions of rest-frame UV photons with \HI{} atoms in gaseous clouds along the sightline, have been increasingly employed to investigate the high-redshift LSS. 
Differently from galaxies, which identify discrete points in redshift space, the \Lyaf{} is a valuable proxy for the continuous matter density field along the line-of-sight at $2\lesssim z \lesssim 4$, allowing to use multiple quasar sightlines to reconstruct the three-dimensional matter density field \citep{Nusser1999,Pichon2001,Gallerani2011,Kitaura2012,Horowitz2019,Huang2020,Porqueres2020}. Detailed cosmological simulations \citep[e.g. ][]{Cen1994,Hernquist1996,MiraldaEscude1996,Zhang1998,Gnedin1998,Dave1999,Machacek1999,Machacek2000,Regan2007,Norman2009,Lukic2015,Nagamine21} indicate that the forest traces preferentially the underdense and mildly overdense ($\delta\lesssim 1-10$) regions, whereas the absorption lines rapidly saturate in corresponding high-density gaseous clouds, probing therefore a complementary density regime to that traced by galaxies.

The \Lyaf{} has become a valuable cosmological probe, exploiting the Baryon Oscillation Spectroscopic Survey \citep[BOSS,][]{Eisenstein2011,Dawson2013} $\sim160,000$ quasars dataset to measure the baryon acoustic oscillations (BAO) \citep{Slosar2011,Busca2013,Slosar2013,Kirkby2013,FontRibera2013,FontRibera2014b,Delubac2015,Bautista2017,Bourboux2017,Bourboux2020} and recently even proposed to constrain the growth rate \citep{Cuceu2021}. As surveys, such as the Dark Energy Spectroscopic Instrument (DESI) survey \citep{Levi2019TheDESI}, get ready to improve on cosmological measurements, the modelling of the \Lyaf{} needs to improve as well.

While the \Lyaf{} is a linear tracer of the underlying density field on large cosmological scales ($k< 0.1\,h\,\rm{Mpc}^{-1}$), on smaller scales non-linear corrections become unavoidably necessary \citep[e.g.][]{Chen2021}, such that modelling the absorption field at high-resolution turns out to be a highly non-trivial task. Additional effects such as \HI{} self-shielding \citep{FontRibera2012b}, redshift space distorsions (RSD hereafter, see   \citet[][]{Kaiser1987,Hamilton1998} and in particular for the \Lyaf{} \citet[]{Seljak2012}), and thermal broadening \citep{Cieplak2016}, among others, complicate even further the overall analytical framework. 
The large degree of complexity accounted for by these features traditionally requires resorting to full hydrodynamic simulations, at the expense of huge time and computational resources, making it practically unfeasible to produce the large amount of independent cosmic volumes necessary to perform cosmological analysis and exploit in full power the unprecedent incoming datasets from future surveys.

To overcome this problem, a customary approach has been to generate low-resolution \Lyaf{} cosmological volumes using scaling relations \citep[Fluctuating Gunn-Peterson Approximation, FGPA hereafter,][]{Rauch1998,Croft1998,Weinberg1999} applied to independent dark matter (DM) fields, obtained from DM-only N-body simulations \citep{Meiksin2001,Viel2002,Slosar2009,White2010,Rorai2013,Lee2014}, approximated gravity solvers \citep{Horowitz2019} (e.g. \texttt{FastPM} \citep{2016MNRAS.463.2273F}, \texttt{COLA} \citep{Tassev2013}), Gaussian random fields \citep{LeGoff2011,FontRibera2012,Bautista2015,Farr2020} and rank-ordering transformations from linear theory to non-linear distributions obtained from a hydrodynamic simulation \citep{Viel2002,Greig2011}. Although the FGPA succeeds in roughly accounting for the spatial correlations on large scales, in the case of $N$-body simulations it is found to lose accuracy at $k>0.1\,h\,\rm{Mpc^{-1}}$ (as will be reported in more detail in \S\ref{sec:bam} and \ref{sec:pipeline}). 

Approaches relying on Gaussian random fields need to be complemented with an input power-spectrum to impose a priori the two-point target summary statistics. However, they do not guarantee to accurately  reproduce the realistic three-dimensional cosmic-web matter distribution, as would be obtained from a simulation, or a detailed analytical structure formation model. In particular, \cite{FontRibera2012} assumes a lognormal model for the \Lyaf{} optical depth $\tau$ distribution, obtained applying a non-linear transform to the Gaussian random field and fitting the free parameters to match the mean transmitted flux to observations. This prescription holds in first approximation, although it does not ensure that the flux distribution accurately matches the observed one. A similar approach adopted by \cite{LeGoff2011}, combining the lognormal model with the FGPA, features substantial deviations in the $1$D power spectrum at $k\gtrsim 0.4\,h\,\rm{Mpc}^{-1}$ \citep{LeGoff2011}. 

The accuracy of the \Lyaf{} modelling methods is expected to improve when using (fast) $N$-body codes rather than approximated gravity solvers, which fail to model the evolution of cosmic structures beyond the mildly non-linear regime.
However, given the high redshift of the data and when the studies are restricted to large scales of say, above $\sim5\,h^{-1}\rm{Mpc}$, fast methods correcting for shell-crossing may be accurate enough \citep[][]{2013MNRAS.435L..78K,2021MNRAS.505.2999T}.

A number of efforts have been dedicated to develop alternative techniques based on matching just the \Lyaf{} probability distribution function (PDF, hereafter) \citep[e.g. the \texttt{LyMAS} code,][]{Peirani2014} and/or the power-spectrum \citep[as e.g. in the \textit{Iteratively Matched Statistics}, IMS,][]{Sorini2016}. However, in such cases the achieved accuracy in the one- and two-point summary statistics (IMS: $4\%$ mean deviations in the PDF and $5\%-7\%$ in the power-spectrum) goes in detriment of the bias relation and bi-spectrum \citep[for a similar situation modelling the halo distribution, see][]{Kitaura2015}. An accurate modelling of the two- and three-point statistics is mandatory to obtain proper covariance matrices that leads to unbiased likelihood analysis of the data \citep[see e.g.,][]{Baumgarten2018}.

Two recent companion papers have proposed to use convolutional neural networks to synthetize hydrodynamic fields, including the Lyman-$\alpha$ forest, starting from the dark matter field, using a deterministic \citep{Harrington2021} or a probabilistic \citep[\texttt{HyPhy,}][and references therein]{Horowitz2021} deep learning approach. Such methods outperform previous efforts, although they still feature significant biases already at the level of reproducing the \Lyaf{} PDF.

In this work we present the latest application of the \textit{Bias Assignment Method} \citep[][\texttt{BAM} hereafter]{Balaguera2018} to map the \Lyaf{} field in three dimensions from a reference cosmological hydrodynamic simulation in order to establish a pipeline and yield an accurate (and fast) alternative to the aforementioned methods for the generations of hydro-simulations. 

In particular, we show that accurate reconstructions of the \Lyaf{} cannot be obtained directly from the dark matter field in our framework, but require using a hierarchical domain specific machine learning  approach  which involves the reconstruction of ionized (\HII{}) and neutral (\HI{}) gas density fields first, as proposed in \cite{Sinigaglia2020} (HydroBAM-I hereafter, equivalent to BAM-V). Subsequently, we use such fields to model for the first time the non-trivial mapping of Ly$\alpha$ fluxes, thereby including the dependence on baryon effects, such as cooling and feedback. We introduce here also a novel framework for the treatment of redshift-space distortion, not included in HydroBAM-I. In summary, while HydroBAM-I introduces the general hierarchical bias showing reconstructions of \HII{}, \HI{} and $T$ fields in real space, in this work we extend our framework to the complex treatment of \Lyaf{}, both in real and in redshift space. In particular, we start by considering the problem of predicting the \Lyaf{} field from the dark matter field and argue that a direct mapping between these two quantities is not sufficient to achieve the desired accuracy in the power spectrum and higher-order summary statistics. Then, we show how the bias hierarchy introduced in HydroBAM-I helps in accomplishing our goal and provide physical insights to explain this performance. Furthermore, we extend the whole framework to redshift space, devising a suitable model for redshift space distortions.

The performance of our method is assessed by means of the deviation of relevant summary statistics (power-  and bi-spectrum in real and redshift space) with respect to the reference simulation. The results suggest that the implementation of our method with hydrodynamic simulations on cosmological volumes will be in position to produce a set of accurate \Lyaf{} mock realizations in a fast accurate way, with direct applications to upcoming surveys such as DESI  and WEAVE \citep[][]{Dalton2012}.

The paper is organized as follows. In \S\ref{sec:bam} we briefly introduce the \texttt{BAM} method, describe the reference cosmological hydrodynamic simulation and summarize the FGPA. In \S\ref{sec:pipeline} we present our pipeline, which is applied to \Lyaf{} reconstructions in real and redshift space in \S\ref{sec:rec_real} and \S\ref{sec:rec_redshift}, respectively. Possible limitations of our method are discussed in \S\ref{sec:bam_lim}. We end with conclusions in \S\ref{sec:conclusions}.

\section{Hydro-BAM} \label{sec:bam}

In this section, we recap first the machine learning method we use in our framework. Secondly, we describe the training data set used to learn the modelling of the \Lyaf{}, in particular a large hydrodynamic simulation.

\subsection{The machine learning method} \label{sec:bam_ml}

We build on the \texttt{BAM} algorithm \citep[][BAM-I and BAM-II hereafter]{Balaguera2018,Balaguera2019}.
As stated in HydroBAM-I, this is a special domain specific machine learning approach which learns the values of a binned isotropic kernel in Fourier-space, minimizing the cost function defined by the Mahalanobis distance of the reconstructed power-spectrum and the ground truth given by the reference simulation. The minimization is achieved through an iterative rejection  Metropolis-Hastings algorithm. 
The kernel preserves the dimensionality of the problem, meaning that it is equivalent to a quadratic matrix in configuration space. Hence, non-local dependencies need to be explicitly provided within the \texttt{BAM} algorithm, otherwise they will only be accounted for isotropically  through the kernel and the anisotropic contributions will be lost\footnote{We know that the relevant non-local dependencies are not isotropic \citep[see e.g.][and references therein]{Kitaura2020}.}. In this sense, \texttt{BAM} represents a physically motivated supervised learning  algorithm. The advantage is that knowing the non-local bias contributions and hierarchical bias relations, we can speed up the learning process using only one (sufficiently large) or a few reference simulations with same primordial power spectrum, setup,  and cosmological parameters, e.g. fixed-amplitude paired simulations \citep[][which have shown that just two fixed-amplitude paired simulations can suppress cosmic variance and achieve the same accuracy in the power spectrum as tens of traditional simulations]{Angulo2016}, thereby increasing the accuracy of the reconstruction, as we will show below. In the latter case, using different
simulations is mainly meant at increasing the effective volume covered by the training set, hence making the bias relations less prone to be affected by cosmic variance. In this sense, it is crucial that such different simulations share the same numerical setup, parameters and parametrizations, as mentioned above, not to introduce incoherencies in the training set.

The resulting kernels and bias relations (the latter understood as the probability $\mathcal{P}(\eta| {\Theta})_{\partial V}$\footnote{We will omit the symbol $\partial V$ hereafter.} of having a given number of dark matter tracers $\eta$ in a cell of volume $\partial V$ conditional to a set $\Theta$ of properties of the dark matter density field, DMDF hereafter in this section, in the same cell can be used to  generate independent realizations of the tracer spatial distribution, when mapped over independent DMDF (evolved from a set of initial conditions generated with the cosmological model and parameters used for the reference simulation). \texttt{BAM} has been shown to be able to generate ensembles of mock catalogs with an accuracy of $1-10\%$ in the two- and three-point statistics of dark matter halos (see e.g. BAM-II), with higher accuracy that previous bias-mapping methods \citep[e.g.][BAM-III]{Pellejero2020} and currently aims at generating the realistic set of galaxy mock catalogs by including intrinsic properties of tracers such as halo masses, velocity concentrations, spins (Balaguera-Antol\'{\i}nez \& Kitaura, in preparation).

Our classification of \texttt{BAM} as a machine learning algorithm relies on the fact that it attempts to solve the problem of assessing the optimal link (bias relation) between the dark matter density field and some tracer field. By optimal we mean the relation yielding the most accurate predictions of one- and two-point statistics of the target tracers field when mapping it from the dark matter field. In particular, \texttt{BAM} learns how to modify, through a kernel, the bias initially measured from a reference simulation acting as training set, applying it, in a second stage, to an independent  and approximated DMDF, in order to predict mock catalogs of the desired target tracers field. As anticipated, in \texttt{BAM} the cost function is defined as the Mahalanobis distance between the reconstructed power-spectrum and the ground truth given by the reference simulation. In this picture, the Metropolis-Hastings algorithm employed in the iterative determination of the kernel, is the optimization step usually embedded in the training phase of a machine learning algorithm, as e.g. gradient descent in a regression problem. However, since we are in the presence of a stochastic mapping between the dark matter field and its tracers field, we cannot rely on a deterministic minimization algorithm based e.g. on gradient computation. Instead, we need to employ an alternative algorithm (the Metropolis-Hastings in our case) which performs optimization by means of Monte-Carlo methods.

In HydroBAM-I, the approach of \texttt{BAM} was extended to the context of hydrodynamic simulations\footnote{This application is dubbed \texttt{Hydro-BAM}, but we will sometimes refer to it throughout the paper as just \texttt{BAM} for shortness.}, exploring the link between gas properties and the underlying DMDF, as well as the scaling relations between different gas properties.
According to our findings, the spatial distribution of \HII{} represents an intermediate stage between the dark matter density field and other baryon quantities, tracing both the large-scale matter distribution, as well as small-scale baryonic processes and the underlying thermodynamic relations.


The \texttt{BAM} procedure can be briefly summarized as follows:
\begin{itemize}
    \item Start with a \texttt{DMDF} $\delta_{\rm dm}(\vec{r})$ interpolated on a mesh with resolution $\partial V$.
    \item For a given dark matter tracer $\eta$, obtain the interpolation on the same mesh as the DMDF. 
    \item Measure the bias $\mathcal{P}(\eta|\{\Theta\})$, where $\{\Theta\}$ denotes a set of properties of the underlying dark matter density field $\delta_{\rm dm}(\vec{r})$ retrieved from  the reference simulation. This bias is obtained from the assessment of the joint distribution $\mathcal{P}(\eta,\{\Theta\})$, which is represented by a multi-dimensional histogram consisting on $N_{\eta}$ bins of the target tracer property and $N_{\Theta}$ properties of the DMDF.
    \item Sample a new variable $\tilde{\eta}$ at each spatial cell (with DM properties $\{\tilde{\Theta}\}$) as 
    \begin{equation}\label{eq:sam_raw}
        \tilde{\eta}\curvearrowleft \mathcal{P}(\tilde{\eta}=\eta|\{\Theta\}=\{\tilde{\Theta}\}) 
        \end{equation}
    \item Measure the power-spectrum  of this quantity and compare with the same statistics from the reference.
\end{itemize}

In general, the first sampling of the tracer field performed according to the bias relation learnt from the simulation results in biased reconstructions of the power-spectrum $P_{\tilde{\eta}}(k)$ of the variable $\tilde{\eta}$ with respect to the reference simulation statistics $P_{\eta}(k)$.
Such deviation is accounted for and corrected iteratively by means of a convolution of the dark matter density field with an isotropic kernel $\mathcal{K}(k=|\vec{k}|)$, selected with a Metropolis-Hastings algorithm and updated throughout the iterations. Briefly, the process to obtain the kernel starts with the definition of the ratio $T_{0}(k)\equiv P^{i=0}_{\tilde{\eta}}(k)/P_{\eta}(k)$, where $P^{i=0}_{\tilde{\eta}}(k)$ is the power-spectrum  obtained from the sampling described by Eq.~(\ref{eq:sam_raw}). At each iteration $i$ (producing a new estimate of $P^{i}_{\tilde{\eta}}(k)$), a value of the ratio $T_{i}(k)$ is computed. At each wavenumber $k$, it is subject to a Metropolis-Hasting (MH) rejection algorithm\footnote{This is done by computing a transition probability ${\rm min}(1, {\rm exp}(\mathcal{H}^{2}_{0j}-\mathcal{H}^{2}_{1j}))$ with $\mathcal{H}_{ij}=(P_{\rm ref}-P^{i}_{\tilde{\eta}})/\sigma$, $\sigma=\sigma(k)$ the (Gaussian) variance associated to the reference power-spectrum.}. The outcome of the MH sampling process defines a sets of weights $w_{i}(k)=T_{i}(k)$ if the current value of $T_{i}(k)$ is accepted, or $w_{i}(k)=1$ otherwise. The \texttt{BAM} kernel is then defined as $\mathcal{K}(k)\equiv \prod_{j=0}^{j=i}w_{j}(k)$ such that, under successive convolutions with the DMDF, the limit $\lim_{n\to \infty} P^{(n)}_{\tilde{\eta}}(k)/P_{\eta}(k)\to 1$ (where $n$ is the number of iterations) is induced.

The aim of this operation is to transform the underlying DMDF such that, once the sampling of the tracer-bias over the DMDF is performed, we obtain a tracer field with the reference power-spectrum. This is achieved in BAM with typically $\lesssim 1\%$ deviations\footnote{Throughout this work, deviations (sometimes referred to as residuals) between two quantities $A(x)$ and $B(x)$ are computed as $\sum_{i}|A(x_{i})/B(x_{i})-1|/N_{x}$ where the sum is done over all probed (bins of) values of the variable $x$ and $N_{x}$ is the number of $x$ values.} after $\sim200$ iterations. The convergence is nevertheless sensitive to i) the cross-correlation between the DMDF and the tracer field, ii) the properties of the dark matter density field included in the bias model, and iii) the way these properties are explored (i.e., binning strategy). A weak fulfillment of any of these conditions may potentially undermine the success of the whole procedure (see also HydroBAM-I).

Furthermore, it is desirable that $\mathcal{K}(k)<1 ~\forall k$, since otherwise it is equivalent to a deconvolution in a similar way to correcting for aliasing introduced by a mass assignment scheme \citep[][]{2005ApJ...620..559J}. It is known that deconvolving a signal in the presence of noise can  enhance the noise contribution\footnote{Let us suppose that we aim at recovering a signal $s$ from some simple data model expressed as: $d={\rm R}s+\epsilon$, where R is a response function and $\epsilon$ some noise contribution. If R is modelled as a convolution, then reconstructing the signal with  $s_{\rm rec}={\rm R}^{-1}d=s+{\rm R}^{-1}\epsilon$ will yield reasonable results as long as the additive noise term remains small.} \citep[see e.g.][and references therein]{2008MNRAS.389..497K}. In our case, the bias relations are subject to the particular realisation and include stochastic components, which  are also affected by the mathematical representation such as  the binning, hence, introducing a noise component.   In the absence of a prior optimal non-linear transformation which regularises the kernel to values $\mathcal{K}<1$, we allow the kernel to go beyond $1$, which is theoretically sub-optimal as might in principle enhance the noise contribution. As we show in \S \ref{sec:rec_real}, however, from a practical point of view it yields already very good results. A further improvement on this is left for future work when smaller scales will be investigated in more detail. 

Since \texttt{BAM} is based on a direct measurement of the bias relation (described as a multi-dimensional histogram\footnote{See BAM-II for the details.}), devising an efficient way to represent the different properties is a key ingredient in our analysis. 
This is done by means of a suitable scaling and binning, aiming at efficiently extracting the information encoded in  the joint probability distribution describing the correlations between physical quantities. The total number of bins used to model the properties in $\{\Theta\}$ is $N_{\rm bins,tot}=\prod_{k=1}^n \theta_k$, where $\theta_k$ is the number of bins used to model property $k\in\{\Theta\}$, and therefore rapidly grows with $n$. 
For this reason, we need a sufficiently large volume or several simulations to avoid over-fitting. 

We refer the reader to \S4.4 of HydroBAM-I for a preliminary assessment of overfitting in the context of \texttt{Hydro-BAM} when learning from small data sets. We further discuss this subject in forthcoming sections. 


In comparison to other machine learning approaches, adopting a deep learning neutral networks perspective \citep[e.g.,][]{Harrington2021,Horowitz2021}, \texttt{Hydro-BAM} relies on a simpler architecture with no hidden layers. As mentioned above in this Section, the crucial physical dependencies to map the desired dark matter tracer field need to be explicitly specified in the bias model implemented in the code. The kernel, indeed, modifies iteratively the dark matter field in order to  try to compensate the lack of fundamental physical information in the bias formulation. In particular, previous applications (BAM-I to BAM-V) have clearly highlighted that including non-local anisotropic contributions in the bias accounts for significant improvement of the tracer mapping.


\subsection{The training data set}

The training data set is determined by a reference cosmological hydrodynamic simulation. This simulation has been obtained using the cosmological smoothed-particle hydrodynamics (SPH) code \texttt{GADGET3-OSAKA} \citep{Aoyama2018, Shimizu2019, Nagamine21}\footnote{\texttt{GADGET3-OSAKA} is a modified version of \texttt{GADGET-3}, an updated version of the popular $N$-body/SPH code \texttt{GADGET-2} \citep{Springel2005}}, with comoving volume $V=(100\,h^{-1}\text{Mpc})^3$, $N=2\times512^3$ particles of mass $m_{\rm DM}=5.38\times10^8\,h^{-1}\text{M}_\odot$ for DM particles and $m_{\rm gas}=1.0\times 10^8\, h^{-1}\text{M}_\odot$ for gas particles, gravitational softening length $\epsilon_g = 7.8 \,h^{-1}$ kpc (comoving) and allowing the baryonic smoothing length to become as small as $0.1\,\epsilon_g$. As a result, the minimum baryonic smoothing at $z=2$ is about physical $260\,h^{-1}$ pc, which is sufficient to resolve the \HI{} distribution in the circumgalactic medium. Some basic statistics of Ly$\alpha$ forest have already been discussed in \citet{Nagamine21}. 
The simulation includes detailed models for supernova and stellar feedback \citep{Shimizu2019}, photo-heating and photo-ionization under a uniform UV background \cite[][]{Haardt2012}, radiative cooling, chemical evolution of atomic and molecular species by \texttt{Grackle} \citep[][]{Smith2017}, among others. The simulation is evolved from initial conditions generated at $z=99$ using \texttt{MUSIC} \citep{Hahn2011}, adopting \cite{Planck2016} cosmological parameters.   

For the purpose of this work, we consider the output at $z=2$ of the simulation, interpolating dark matter density, \HII{} density, \HI{} number density and \HI{} optical depth onto a $128^3$ cubic mesh using a CIC mass assignment scheme (MAS hereafter). In this setup, the physical cell-volume corresponds to $\partial V \sim (0.78\,h^{-1}\rm{Mpc})^{3}$ and the Nyquist frequency is $k_{N}=4.02\,h\,\rm{Mpc}^{-1}$. 
\begin{figure*}
\centering
\includegraphics[width=18cm]{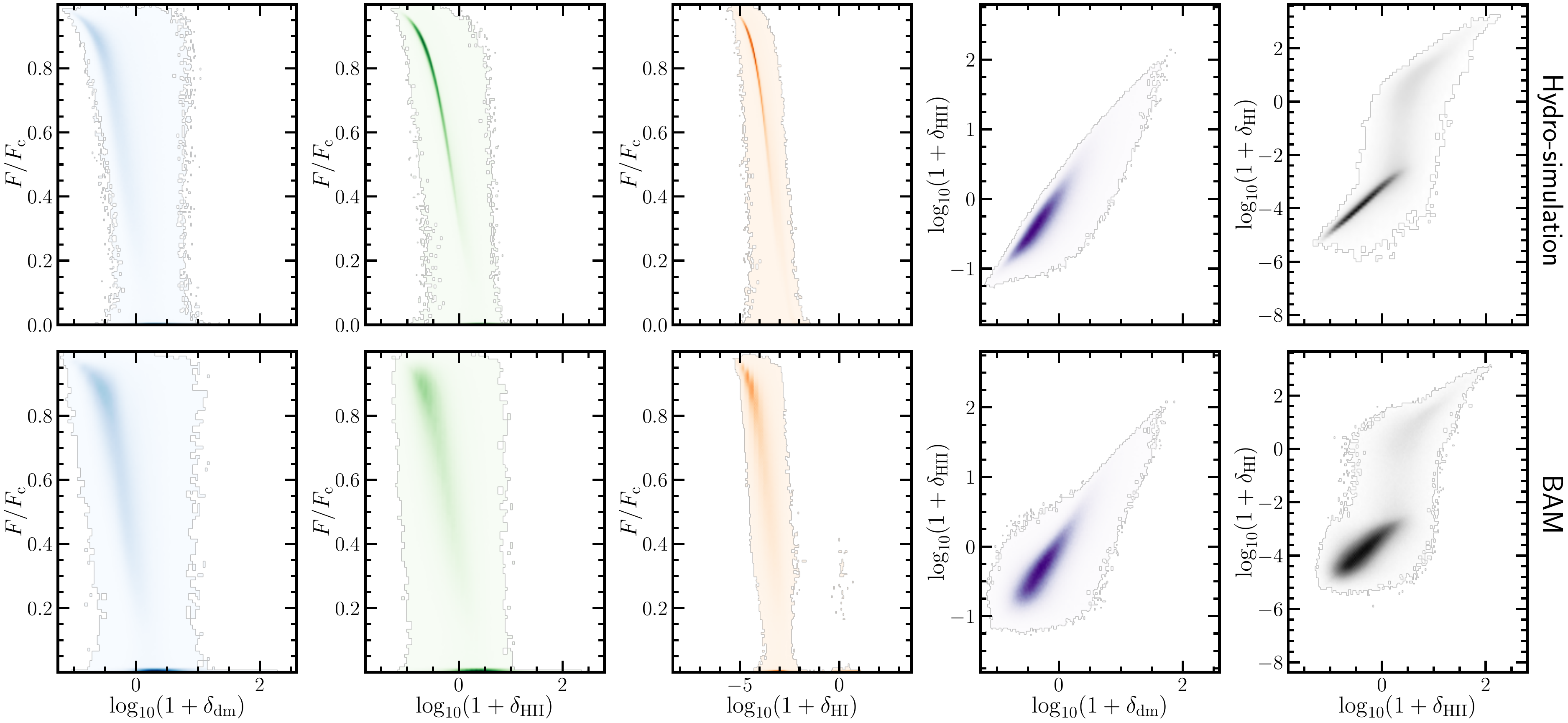}
\caption{\small{Phase diagrams showing the scaling relations of the transmitted flux $F/F_{c}$ as a function of different properties of the gas and dark matter, from the reference simulation (top row) and the \texttt{BAM} reconstructions (bottom row). The color intensity indicates the height of the joint probability distribution between each pair of properties. White regions correspond to empty two-dimensional bins, i.e. to zero-probability regions. The smoothing is achieved through a bilinear interpolation over a $100\times100$ bins grid.}}
\label{fig:phase_diagram}
\end{figure*}

The measurements of power- and bi-spectra are performed within spherical averages in Fourier space of width equal to the fundamental mode, with proper corrections for the MAS implemented \citep[through its Fourier transform, see e.g.][]{1981csup.book.....H} and for their (Poisson) shot-noise.

We refer the reader to \S2 of HydroBAM-I \citep[and references therein, e.g.,][]{Nagamine21} for a detailed description of the reference simulation and the properties of the dark matter and baryon fields taken into consideration.

A major novel aspect of this work with respect to HydroBAM-I is the treatment of \HI{} optical depth $\tau$ and the corresponding \Lyaf{} field \footnote{The notation $F=\exp(-\tau)$ for the \Lyaf{} flux field actually refers to the transmitted flux $F/F_{\rm c}$, i.e. to the quasar spectrum normalized to its continuum $F_{\rm c}$. We will omit hereafter the normalization and denote the transmitted flux just as $F$, unless specified otherwise.}$F=\exp(-\tau)$. In this context, we do not adopt the FGPA, but rather obtain the \HI{} opacity by means of a line-of-sight integration. The optical depth is computed as \citep{Nagamine21}: 
\begin{equation}\label{eq:tau}
    \tau = \frac{\pi e^2}{m_e c} \sum_j f \, \phi(x-x_j) n_{\rm HI}(x_j) \Delta l,
\end{equation}
where $e$, $m_e$, $c$, $n_{\rm HI}$, $f$, $x_j$, $\Delta l$ denote respectively the electron charge, electron mass, speed of light in vacuum, \HI{} number density, the absorption oscillator strength, the line-of-sight coordinate of the $j$-th cell and the physical cell size. The Voigt-line profile $\phi(x)$ in Eq.~(\ref{eq:tau}) is provided by the fitting formula of \cite{Tasitsiomi2006}. Where necessary, relevant quantities (e.g. \HI{} number density) are previously interpolated on the mesh according to the SPH kernel of the simulation. Coordinates $x_j$ of cells along the line-of-sight refer to the outcome of interpolation of particles based either on their positions in real space $r_j$  or redshift-space $s_j=r_j + v^{\rm los}_j/aH$, where $v^{\rm los}_j$, $a$ and $H$ are the $j$-th particle velocity component along the line-of-sight, the scale factor and the Hubble parameter at $z=2$, respectively.

\begin{figure*}
\centering
\includegraphics[width=18.5cm]{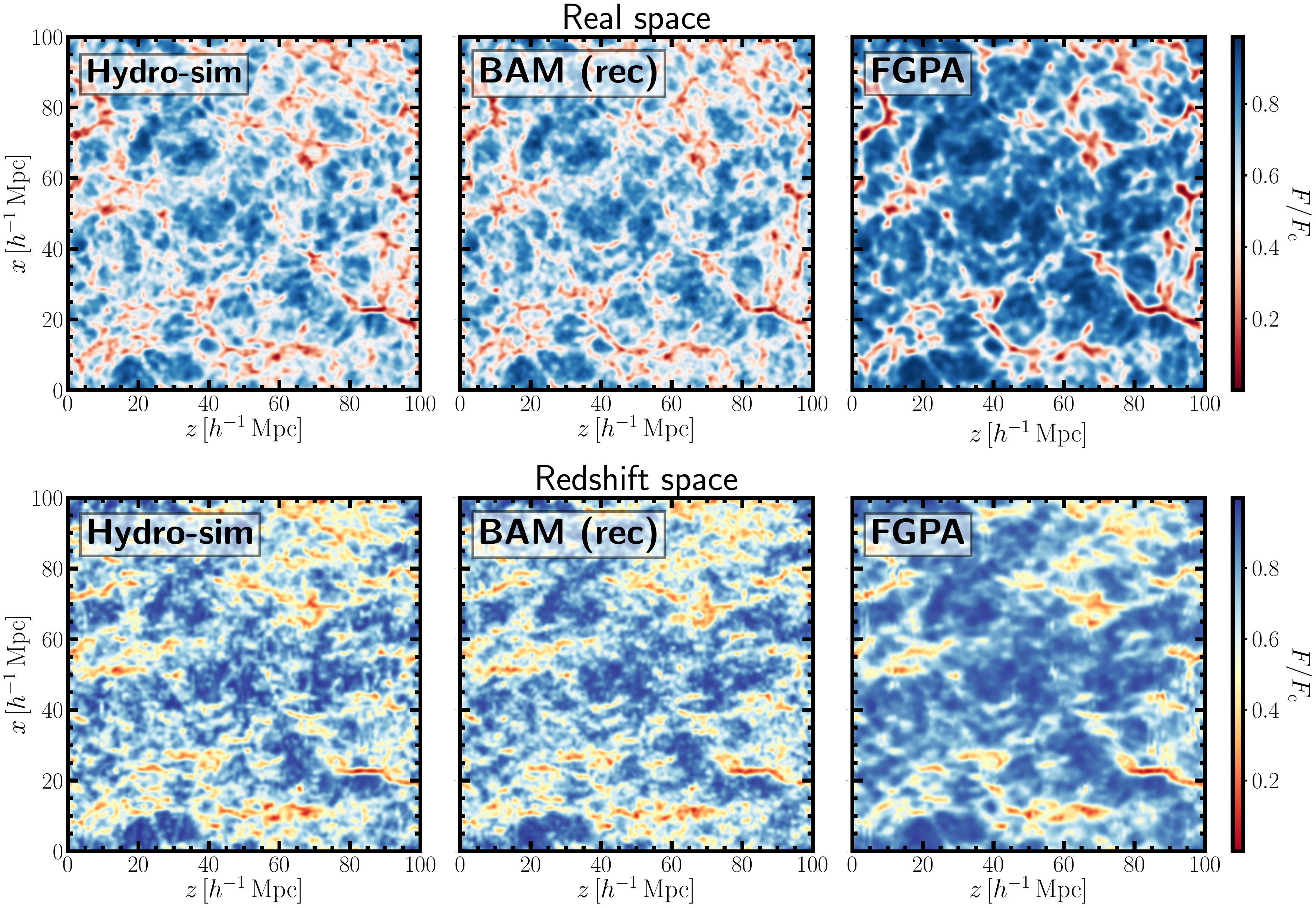}
\caption{\small{\Lyaf{} slices through the simulation box in real (top row) and redshift space (bottom row). From left to right: a slice from reference simulation, from \texttt{BAM} reconstruction using \HII{} and \HI{} previously calibrated within \texttt{BAM}, and from FGPA box. These panels evidence how \texttt{BAM} is optimal at reproducing both large and small-scale structures, in contrast to the reconstructions performed with the FGPA, where small-scale structures are evidently suppressed.}}
\label{fig:slices}
\end{figure*}

\subsection{Fluctuating Gunn-Peterson Approximation}

Hydrodynamic simulations \citep[see e.g.,][]{Lukic2015} show that a considerable fraction ($>90\%$ in volume and $>50\%$ in mass) of the gas probed by the \Lyaf{}, found in regions with mildly-linear ($\delta_{\rm dm}<10$) regime, is in a diffuse state not subject to shock processes. In these conditions, the gas density $\rho_{\rm gas}$ and temperature $T_{\rm gas}$ are found to display a tight scaling relation characterized by a power-law relation $T_{\rm gas}=T_0(\rho_{\rm gas}/\bar{\rho}_{\rm gas})^\alpha$ (with $\bar{\rho}_{\rm gas}$ being the average gas density), whose parameters (amplitude and slope) depend on the reionization history and on the spectral slope of the UV background. These vary commonly within the ranges $4000\,\rm{K}\lesssim T_{0} \lesssim 10^{3}\,\rm{K}$ and $0.3\lesssim\alpha\lesssim 0.6$ \citep[see e.g.,][]{HuiGnedin1997}. Assuming photo-ionization equilibrium, $\tau\propto n_{\rm HI}$ and $n_{\rm HI}\propto \rho^2 T_{\rm gas}^{-0.7}/\Gamma_{\rm HI}$, where $n_{\rm HI}$ and $\Gamma_{\rm HI}$ are the \HI{} number density  photo-ionization rate and $\tau$ denotes the optical depth, respectively, it is possible to express the latter as 
\begin{equation}\label{fgpa}
\tau = A(1+\delta_{\rm gas})^\beta ,
\end{equation}
which represents the FGPA. In this expression, $\beta=2-0.7, \alpha\sim1.6$, and $A$ is a constant which depends on redshift and on the details of the hydrodynamics \citep[see e.g.][]{Weinberg1999}.  Provided that dark matter and gas densities are highly correlated in the cool low-density regions (where the Ly$\alpha$ absorption mostly takes place) $\delta_{\rm gas}$ can be replaced with $\delta_{\rm dm}$ in Eq.~(\ref{fgpa}). Furthermore, in some cases an additional Gaussian smoothing is applied to transform the dark matter field into a pseudo-baryon density field, accounting for the fact that the gas density field tends to be smoother than dark matter density in correspondence of low-density dark matter peaks \citep{Gnedin1998,Bryan1999}.

In this work we compute the FGPA based on our reference simulation and compare it against the results obtained with \texttt{BAM}. To this end, we use the dark matter field at our fiducial resolution ($0.78\,h^{-1}\,\rm{Mpc}$ with a mesh of $128^3$ cells), comparable to that used for cosmological studies, but coarser than typical smoothing lengths employed in detailed studies based on hydrodynamic simulations \citep[e.g., $228\,\rm{kpc}$ in][]{Sorini2016}. We do not apply a Gaussian smoothing. Following relevant works \citep[e.g.,][]{Weinberg1999, Viel2002,Seljak2012,Rorai2013,Lukic2015,Cieplak2016,Horowitz2019}, we use the value $\beta=1.6$, and $A\sim 0.44$, the latter obtained from a linear fit in the $\log_{10}(\tau)-\log_{10}(\delta_{\rm dm})$ plane. This ensures that the resulting scaling relation is consistent with that observed in the reference simulation.

\section{The calibration pipeline} \label{sec:pipeline}

We introduce here a change of paradigm with respect to previous studies of the \Lyaf{} in which the relation of the optical depth is expressed as a function of the dark matter field only \citep[e.g.,][]{McDonald2000,McDonald2003,Seljak2012,Wang2015,ArinyoPrats2015,Cieplak2016,Givans2020,Chen2021}.
Instead, we establish a hierarchy of bias relations starting from a dark matter field,  continuing over the ionised and the neutral gas, to the flux directly.
We circumvent hereby the problem of translating optical depths into fluxes with coarse resolutions. These operations are known to introduce biases, especially taking redshift space distortions into account. In particular, the commutation of non-linear operations (as are down-sampling from the native resolution of the simulation to the resolution adopted in this work, and the mapping from optical depth to \Lyaf{} fluxes $F=\exp(-\tau)$) are known to introduce velocity bias contributions in redshift-space \Lyaf{} representations on a mesh \citep[][]{Seljak2012}. We have explicitly checked with numerical experiments that at the volume and resolution we are working at, this bias is negligible and hence commuting the two aforementioned operations does not imply a problem. This fact offers us the opportunity of applying the $F=\exp(-\tau)$ transform at $128^3$ resolution, thereby yielding directly a reconstruction $F'$ of the final target quantity $F$ and ensuring that our procedure does not introduce further biases. This may have not been the case if we had had a reconstruction $\tau'$ of the optical depth $\tau$ as final step of our hierarchical method, and then computed fluxes as $F'=\exp(-\tau')$. In fact, in that case our procedure would not ensure that the summary statistics of $F'$ reproduces the reference one within the required few percent accuracy.

We construct first a proxy of the gas distribution which is fast to compute. To this end we take the dark matter field from the reference simulation down-sampled to the chosen resolution. A large number of fast gravity solvers are now available achieving that accuracy at those high redshifts ($z=2$) on Mpc scales \citep[][]{2013MNRAS.435L..78K,Tassev2013,2016MNRAS.463.2273F,2021MNRAS.505.2999T}.   Then we make a suitable non-linear transform of that field which improves the bias mapping relation to the targeted tracer distribution.  In particular, we use $\log_{10}(1+X)$ \citep[see][for a motivation]{2009ApJ...698L..90N,2012ApJ...745...17F,2012MNRAS.425.2443K,2012MNRAS.425.2422K}, with $X$ being $\delta_{\rm dm}$, $\delta_{\rm \HII{}}$ and $\delta_{\rm \HI{}}$ \citep[although other kind of transforms can be  considered,][]{Kitaura2020,Sinigaglia2020}.

The key aspect of starting from the dark matter field is that gravitational evolution has shifted the matter perturbations forming a cosmic web which correlates with the gas distribution. In fact, as we showed in HydroBAM-I the correlation between the ionised gas (\HII{}) and the dark matter field is very high.
Despite of this, the bias relation is highly non-linear and non-local. Hence, in a second step, we apply our machine learning algorithm to extract the corresponding kernel and multidimensional bias relation. 

We must stress that the \texttt{BAM} algorithm learns the statistical relations between two fields which can be applied to any spatial configuration.  
One could, in principle, learn the different relations between the dark matter and the baryonic quantities, such as \HII{} or \HI{},  independently. Although, the statistical relations would be individually correct, the thermodynamical equilibrium relations among the various baryonic quantities would not be satisfied, as they would have been obtained ignoring each other.

For this reason, the step in which the first baryonic quantity is sampled fixes the choice of the overall gas distribution.
The hierarchy of bias relations continues with sampling \HI{} from \HII{}. 
We could continue with the temperature of the gas, but our studies have shown that it is not necessary to obtain the \Lyaf{} fluxes. In particular, introducing the dependence on the temperature does not yield substantial improvements in the flux power spectrum calibration, with respect to our fiducial model based just on \HII{} and \HI{}. Therefore, we believe the temperature does not account for crucial missing physical dependencies at our mesh resolution. We will consider potential extensions to the temperature of our model in future works, investigating the accuracy of \Lyaf{} reconstruction towards smaller scales than the ones probed in this paper.

\begin{figure*}
\centering
\includegraphics[width=18cm]{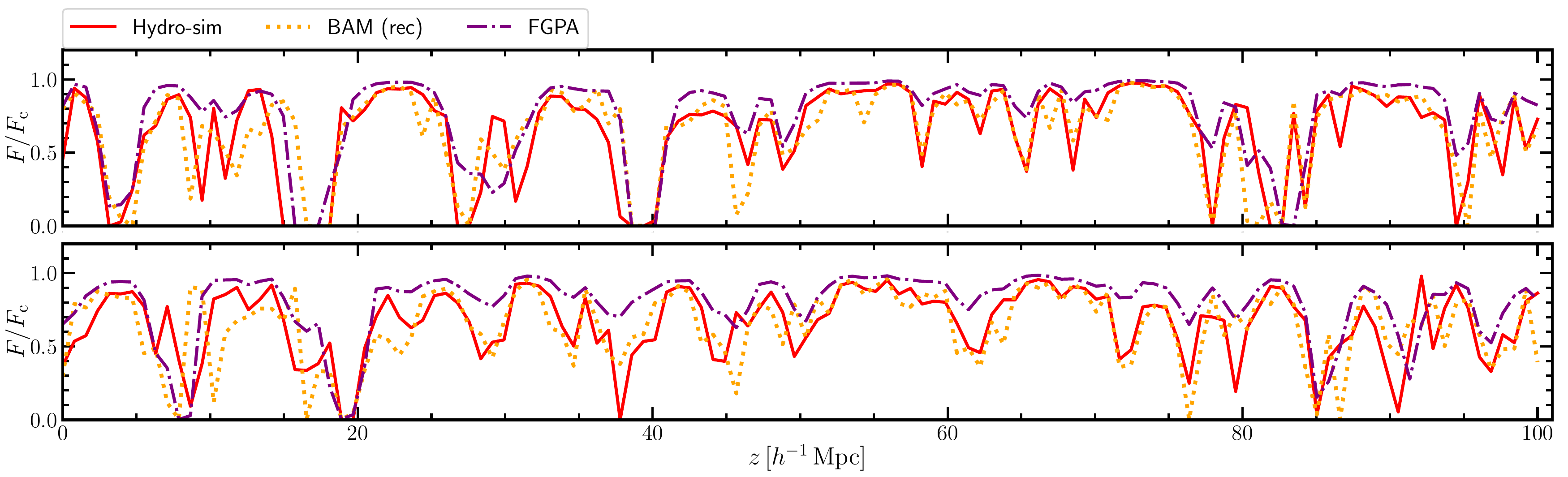}
\caption{\small{Transmitted flux $F/F_{\rm c}$ skewers along the line-of-sight extracted from the simulation box at $2$ random positions. The \Lyaf{} pseudo-spectra reconstructed by \texttt{BAM} using \HII{} and \HI{} fields sampled by \texttt{BAM} (yellow dotted) are found to qualitatively reproduce the reference skewers (red solid) more accurately than the FGPA (purple dash-dotted). The spectral resolution is $l\sim0.78\,h^{-1}\rm{Mpc}$.}}
\label{fig:skewers}
\end{figure*}

The first part of our procedure is entirely performed in real space, whereas the \Lyaf{} mocks are obtained in redshift space. To accomplish this goal we apply \texttt{BAM} in real space to obtain \HII{} and \HI{} densities, and then map these reconstructed quantities in redshift space. This last step requires an adequate modelling of RSD, which will be discussed in \S\ref{sec:rsd}.

The full calibration strategy can be summarized in the following five steps:
\begin{enumerate}
    \item $\tilde{\delta}_{\rm HII}(\vec{r})\curvearrowleft \mathcal{P}\left(\delta_{\rm HII}(\vec{r})|\{\Theta(\delta_{\rm dm}(\vec{r}) \otimes \mathcal{K}_1)\}\right) \equiv \mathcal{B}_1(\vec{r})$
    \item $\tilde{\delta}_{\rm HI}(\vec{r})\curvearrowleft \mathcal{P}(\delta_{\rm HI}(\vec{r})|\{\Theta(\tilde{\delta}_{\rm HII}(\vec{r}) \otimes \mathcal{K}_2)\}) \equiv \mathcal{B}_2(\vec{r})$
    \item $\tilde{\delta}_{\rm HII}(\vec{s})=\mathcal{S}\left[\tilde{\delta}_{\rm HII}(\vec{r}), \vec{v}(\vec{r})\right]$ 
    \item $\tilde{\delta}_{\rm HI}(\vec{s})=\mathcal{S}\left[\tilde{\delta}_{\rm HI}(\vec{r}), \vec{v}(\vec{r})\right]$

    \item $\tilde{F}(\vec{s})\curvearrowleft \mathcal{P}(F(\vec{s})|\{\Theta(\tilde{\delta}_{\rm HII}(\vec{s})\otimes \mathcal{K}_3,\tilde{\delta}_{\rm HI}(\vec{s}))\})=\mathcal{B}_3(\vec{s})$
\end{enumerate}
where $\tilde{X}$ denotes the reconstruction of the property $X$ and $\mathcal{S}$ denotes the mapping from real to redshift space. 
That is, in steps 1, 2 and 5 the field $\tilde{X}$ on the left-hand side is randomly-sampled following the stochastic bias relation $\mathcal{B}$ on the right-hand side, expressing the probability of having a certain value of $X$ in a cell conditional to the set of properties $\Theta$, after convolving  with the kernel $\mathcal{K}$ the field upon which $\Theta$ is built.
As anticipated, steps 1-2, presented in details in HydroBAM-I, are carried out in real space, whereas step 5 is performed in redshift space. $\vec{r}$ and $\vec{s}$ stand for Eulerian coordinates in real and redshift space, respectively. In steps 3-4, $\mathcal{S}$ denotes the mapping from real to redshift space, which will be presented in details in \S\ref{sec:rsd}. In summary, \HII{} and \HI{} densities are obtained in real space starting from the dark matter field in steps 1-2, are moved to redshift space in steps 3-4, and the resulting redshift-space \HII{} and \HI{} fields are used to sample fluxes in step 5.

\begin{figure*}
\centering
\includegraphics[width=18cm]{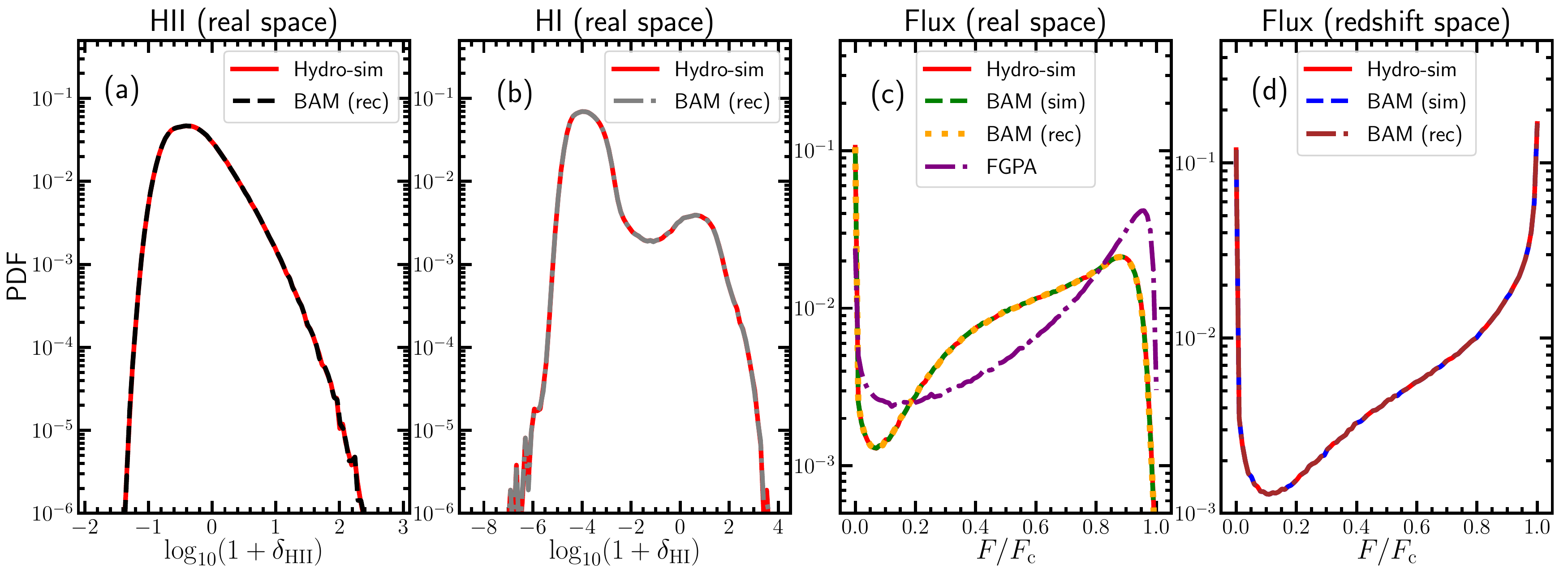}
\caption{\small{Panel (a): distributions of $\log_{10}(1+\delta_{\rm \HII{}})$ in real space from the simulation and as reconstructed by \texttt{BAM} . Panel (b): distributions of $\log_{10}(1+\delta_{\rm \HI{}})$ in real space from the simulation and as reconstructed by \texttt{BAM}. Panel (c): distribution of \Lyaf{} fluxes in real space from the reference simulation, from the FGPA and as reconstructed by \texttt{BAM} based on \HII{} and \HI{} from the simulation and from reconstructions shown in panels (a) and (b). Panel (d): distribution of \Lyaf{} fluxes in redshift space from the reference simulation, as reconstructed by \texttt{BAM} based on \HII{} and \HI{} from the simulation and from reconstructions shown in panels (a) and (b) and mapped in redshift space as explained in \S\ref{sec:rsd}.}}
\label{fig:pdfs}
\end{figure*}

The output of the calibration consists therefore in the sets of multi-dimensional bias relations $\{\mathcal{B}_{\rm i}(\vec{x}=\vec{r},\vec{s})\}$ and kernels $\{\mathcal{K}_{\rm i}\}$, $\rm{i}=1,2,3$. These are then applied to independent dark matter fields to obtain the corresponding \Lyaf{} mock realization, following all the steps depicted above. 

In the following sections, we explore the \Lyaf{} reconstruction in real and redshift space based on baryon density fields and present the results of dedicated \texttt{BAM} runs.

\section{Reconstruction of the Lyman-$\alpha$ forest: real space} \label{sec:rec_real}
In this section we show the performance of the \texttt{Hydro-BAM} approach in real space.

\subsection{The bias model}
We start analysing the reconstruction procedure in real space, i.e., excluding steps 3 and 4 in the hierarchical sampling process introduced in \S\ref{sec:pipeline}. This enables us to assess the accuracy of the method and compare it to the FGPA before additional physical effects (i.e. RSD) are taken into account.

The top panels of Fig.~\ref{fig:phase_diagram} show the phase diagrams (or bias relations) linking the transmitted flux $F$ to other gas properties as well as to the dark matter density in the reference simulation. From this figure it is evident how the relation between the flux and dark matter density shows to be less concentrated (large scatter around the mean relation), in contrast to the scaling relations between the transmitted flux and the \HII{} and \HI{} density fields. This is an indication that the bias model $\mathcal{P}(F|\,\delta_{\rm dm}\otimes\mathcal{K})$ can be sub-optimal for our reconstruction procedure. 
Indeed, when calibrating the bias relation, the resulting power-spectrum of the reconstructed flux field displays $>10\%$ deviations throughout all the probed Fourier modes. 
This result suggests that an arbitrarily complex (i.e, non-linear) $\delta_{\rm dm}-F$ scaling relation is not yet sufficient to capture the full physical correlation between these quantities. Non-local contributions have to be accounted for, which however cannot rely solely on the dark matter field.

The reference hydrodynamic simulation includes a series of baryonic processes (e.g. cooling and feedback by supernovae and active galactic nuclei) relevant to determine the spatial distribution of the gas, especially of its cold neutral phase.

We therefore extend the bias model to include baryons, replacing the dark matter density with gas density fields, i.e,  $\mathcal{P}(F|\{\Theta\})$, where $\{\Theta\}=\{ \delta_{\rm{HII}}\otimes\mathcal{K},\delta_{\rm{HI}}\}$. 

In the case in which \HII{} and \HI{} densities are extracted from the reference simulation, the aforementioned bias model generates reconstructions with percent accuracy in the two-point statistics of the \Lyaf{}, as will be shown in forthcoming sections. The situation in which this bias model uses the \HII{} and \HI{} fields reconstructed with \texttt{BAM} is more subtle. In the latter scenario, we obtain $5\%-10\%$ deviations in the flux power-spectrum on small and intermediate scales ($k\sim 1.0-2.0\,h\,\rm{Mpc}^{-1}$). To reduce these deviations, we add the dependence on the dark matter density with  the model $\{\Theta'\}=\{\delta_{\rm{HII}}\otimes\mathcal{K},\delta_{\rm{HI}},\delta_{\rm{dm}}\}$. This does not further improve the \texttt{BAM} calibration with respect to $\{\Theta\}=\{\delta_{\rm{HII}}\otimes\mathcal{K},\delta_{\rm{HI}}\}$ when using baryons from the simulation. However, in the case of reconstructed \HII{} and \HI{} the explicit dependency with the dark matter density helps to ensure the large-scale clustering signal, allowing for an overall improvement of the reconstruction. We therefore use $\{\Theta\}$ as our fiducial model when \HII{} and \HI{} are read from the reference simulation, while we use $\{\Theta'\}$ when these fields are reconstructed within \texttt{BAM}. 

Adopting two different fiducial models in the two situations poses no conflict if one considers that $\lbrace\Theta'\rbrace$ is just an extension of $\lbrace\Theta\rbrace$, alleviating the slight loss of accuracy discussed above and/or the effect of random noise in the spatial assignment of tracers for reconstructed fields. In the ideal case where \HII{} and \HI{} are extracted from the simulation, we regard $\lbrace\Theta\rbrace$ and $\lbrace\Theta'\rbrace$ to carry an equivalent amount of information of \Lyaf{} clustering, as demonstrated by the numerical tests we mentioned. 

\begin{figure*}
\centering
\includegraphics[width=18cm]{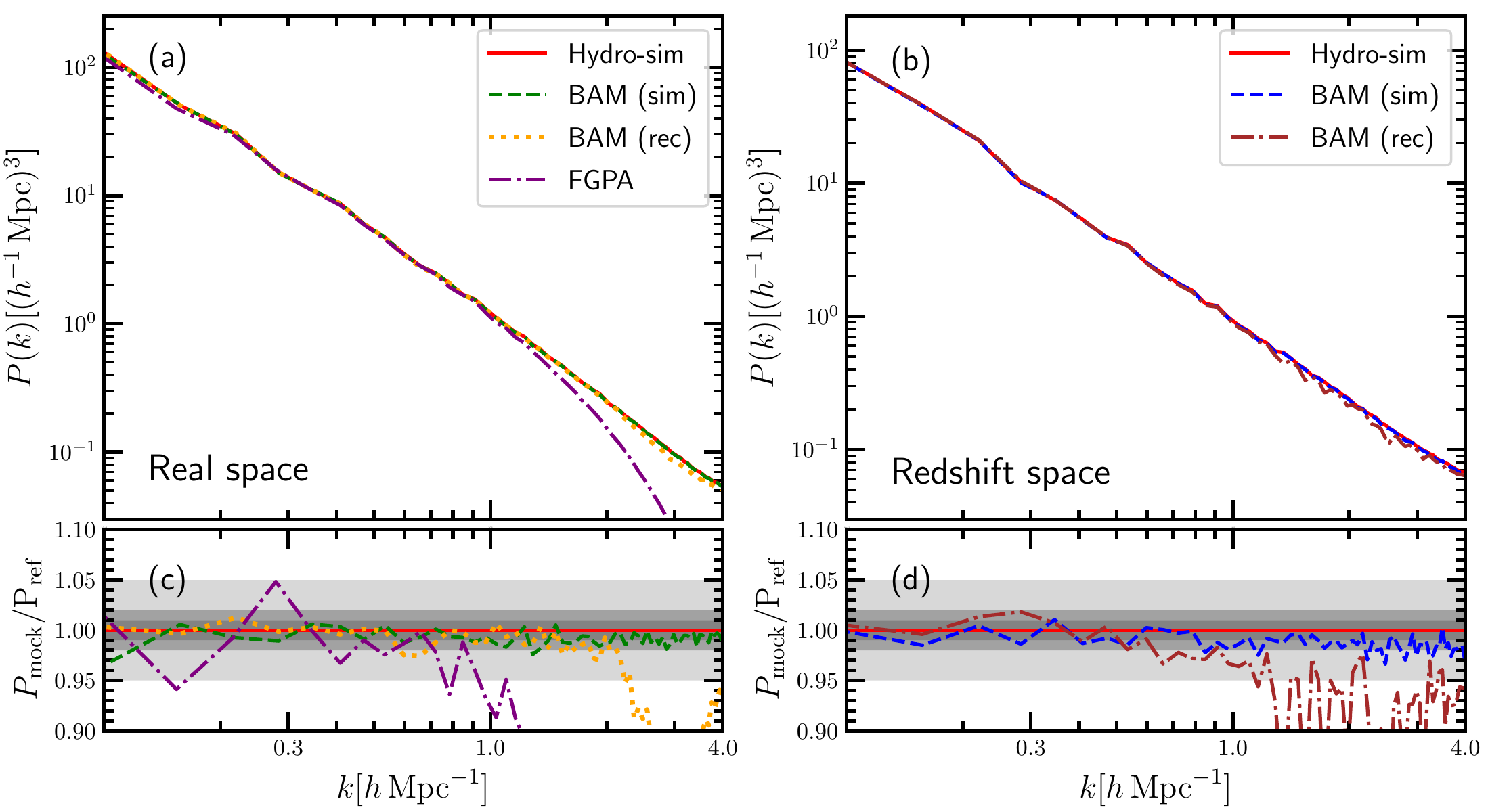}
\caption{\small{Comparison between \Lyaf{} power-spectra in real space (panels (a) and (c)) and redshift space (panels (b) and (d)) of reference simulation, \texttt{BAM} reconstruction using density fields from the reference simulation and \texttt{BAM} reconstruction using \texttt{BAM}-reconstructed density fields. Gray shaded area in panels (c) and (d) indicate $1\%,2\%$ and $5\%$ deviations from the reference simulation.}}
\label{fig:power_spectra}
\end{figure*}

\subsection{Results}
A first visual assessment of the performance of our method in real space is achieved by comparing slices through the simulation box from the reference, the reconstruction by \texttt{BAM} and the FGPA \Lyaf{} fields. This is shown in the top row of Fig.~\ref{fig:slices}, from where we observe that \texttt{BAM} replicates the flux structures, both on large and small scales, in good agreement with the reference. The FGPA, on the other hand, reproduces the large-scale spatial distribution, whereas it tends to suppress small-scale structures and to overestimate the high density regions. 
This can be also observed in Fig.~\ref{fig:skewers}, showing transmitted flux skewers along the line-of-sight extracted at two random positions from the reference simulation, the \texttt{BAM} reconstructions using baryon fields and the FGPA. The origin of the sizable differences between these three cases can be due to the fact that the FGPA accounts only for the $F-\delta_{\rm dm}$ mean relation and for local dependencies on the dark matter field, whereas \texttt{BAM} captures the full non-linear non-local stochastic nature of the tracer-bias.


\begin{figure*}
\vspace{-0.3cm}
\begin{tabular}{c}
\centering
\includegraphics[width=17cm]{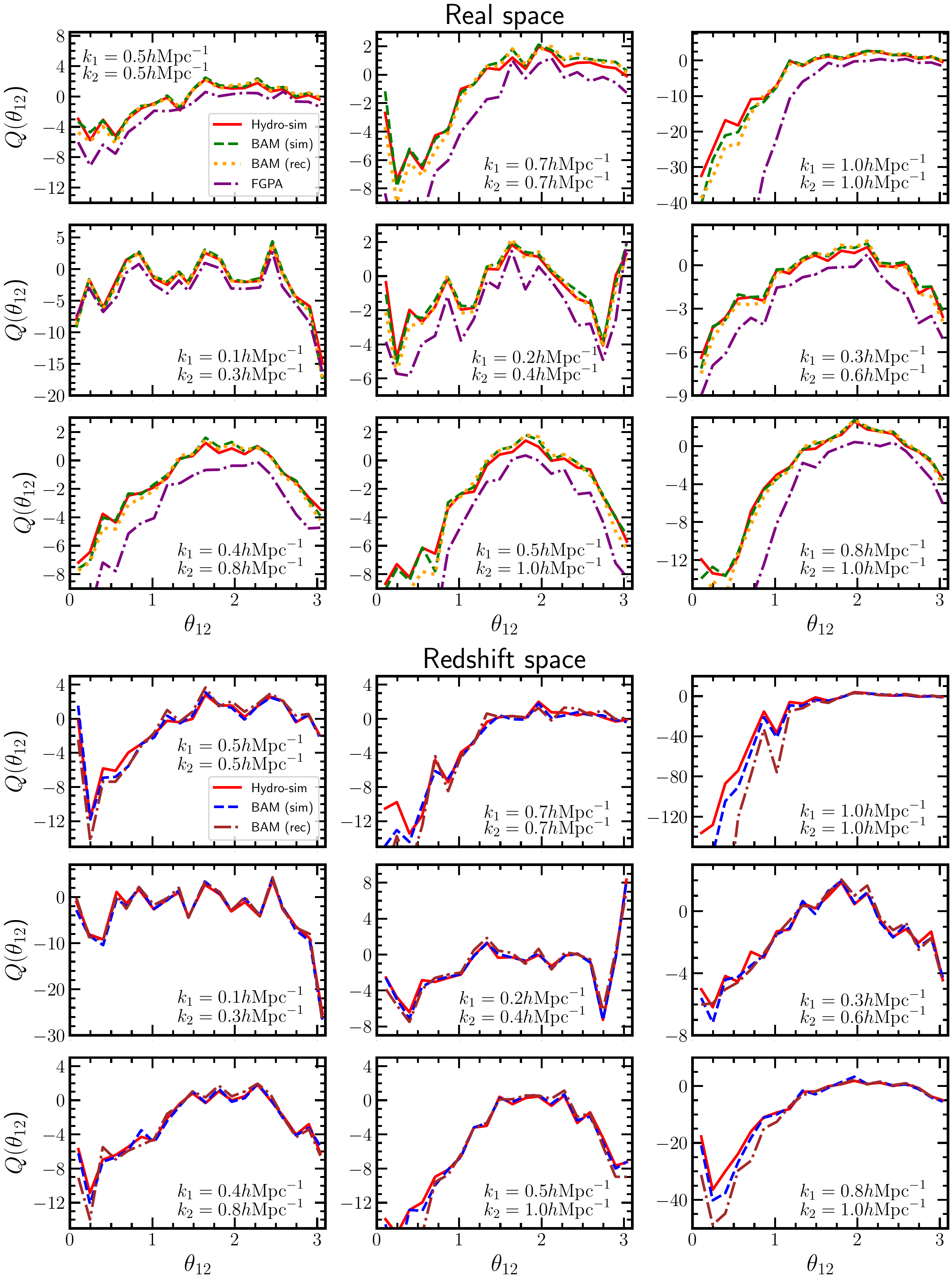}
\end{tabular}
\vspace{-0.4cm}
\caption{\small{Top plots set: real space \Lyaf{} reference bi-spectrum (red solid), bi-spectrum of \texttt{BAM} reconstruction obtained with density fields from the reference simulation (green dashed), FGPA bi-spectrum (purple dash-dotted). Bottom plots set: redshift space \Lyaf{} reference bi-spectrum (red solid), bi-spectrum of \texttt{BAM} reconstruction obtained with density fields from the reference simulation mapped in redshift space (blue dashed) and bi-spectrum of \texttt{BAM} reconstruction obtained with \texttt{BAM}-reconstructed density fields (brown dash-dotted). \texttt{BAM} well-reproduces the reference bi-spectrum up to $k\sim0.8\,h\,\rm{Mpc}^{-1}$ both in real and in redshift space, whereas the FGPA considerably underestimate the reference bi-spectrum at $k>0.1\,h\,\rm{Mpc}^{-1}$.}}
\label{fig:bi-spectra}
\end{figure*}

The quality of the \HI{} and \HII{}  reconstructions  performed by \texttt{BAM} can be firstly assessed through the distribution function of each property in these two scenarios. This is shown in panels (a) and (b) of Fig.~\ref{fig:pdfs}, where we see that \texttt{BAM} reproduces the bi-modality presented in the \HI{} distribution. 
On the other hand, the qualitative differences observed in Fig.~\ref{fig:slices} between FGPA and \texttt{BAM} are represented in  Fig.~\ref{fig:pdfs} panel (c) by an excess of abundance in the low-density ($F\sim1$, i.e., full transmission) regime of the flux distribution function, with a correspondent drop of abundance in high-density regions ($F\sim0$, full absorption). Conversely, \texttt{BAM} reproduces with very high accuracy the flux PDF (panel (d)), as it is designed to match the tracer PDF by construction learning from the reference simulation. 


The panel (a) of Fig.~\ref{fig:power_spectra} and the top set of plots in Fig.~\ref{fig:bi-spectra} show the \texttt{BAM}-reconstructed (green dashed based on reference \HII{} and \HI{}, yellow dash-dotted based on reconstructed \HII{} and \HI{}) and the FGPA (purple dash-dotted) power-spectrum and bi-spectrum\footnote{For the measurements of bi-spectrum we use the code \url{https://github.com/cheng-zhao/bispec} made available by Cheng Zhao.}, compared to the corresponding reference flux field summary statistics (red solid). As can be seen from panel (c) of Fig. ~\ref{fig:power_spectra}, reconstructed power-spectrum based on reconstructed baryon densities (reference baryon densities) average and maximum deviations are respectively $<2\%$ and $\sim 3\%$ up to $k=2.0\,h\,\rm{Mpc}^{-1}$ (up to the Nyquist frequency) with respect to the reference power-spectrum. The FGPA power-spectrum oscillates around the reference one with deviations of  $\sim5\%$ at $k<1.0\,h\,\rm{Mpc}^{-1}$ and is affected by an abrupt drop of power at $k>1.0\,h\,\rm{Mpc}^{-1}$. 
Regarding the bi-spectrum, \texttt{BAM} replicates well the reference bi-spectrum for all the probed configurations, whereas the FGPA bi-spectrum reproduces the qualitative trend but severely underestimates it already at configurations probing scales of $k>0.1\,h\,\rm{Mpc}^{-1}$.

Further experiments considering the \HII{} and \HI{} densities separately, i.e. $\{\Theta\}=\{ \delta_{\rm{HII}}\otimes\mathcal{K}\}$ or $\{\Theta\}=\{\delta_{\rm{HI}}\otimes\mathcal{K}\}$, do not yield the same level of power-spectrum convergence as the case in which the two are used in combination. This reinforces our argument that a comprehensive modelling of both large-scale and small-scale physics is required here. 

Our results can be understood in more detail by comparing panels (g)-(h) and (l)-(m) in Fig.~\ref{fig:phase_diagram}, which show respectively the relations $\mathcal{P}(\delta_{\rm HII}|\,\delta_{\rm dm})$ and $\mathcal{P}(\delta_{\rm HI}|\,\delta_{\rm HII})$ as seen from the simulation and as reconstructed by \texttt{BAM}. Although \texttt{BAM} reproduces the reference simulation remarkably well and manages to replicate the two-phases (hot and cool) thermodynamic structure of the gas over nearly 12 orders of magnitude, the resulting scaling relations turn out to have slightly more scatter than the reference ones, implying that they pose a weaker constraint. This translates into broader bias relations also shown in panels (d) and (f), which need therefore to be complemented with additional information supplied by the dark matter spatial distribution.  

We should bare in mind that the reference hydrodynamic simulation has been run at a considerably  higher resolution (512$^3$ vs 128$^3$). Hence, we cannot expect these relations to be equal, but to be the optimal ones reproducing the summary statistics of the various quantities of interest at a given (lower) resolution (than the reference one). 
The computing time gain at this resolution is already of 7 orders of magnitude as compared to the hydrodynamic simulation. This difference will, however, dramatically increase when going to larger volumes and/or lower mesh resolutions than the ones studied here. In particular, assuming to run \texttt{Hydro-BAM} with fixed number of cores and bias model, the code speed scales approximately as $\mathcal{O}(N_{\rm c})$, with $N_{\rm c}$ the number of cells used in the grid. Therefore, passing from $128^3$ to $64^3$ mesh resolution would imply a gain in computing time of a factor $\sim8$ in the case of the reference simulation used in this paper. Similarly, adopting a simulation with comoving volume $V=(500 \,h^{-1}\,\rm{Mpc})^3$, $N_{\rm p}=2048^3$ particles, interpolated on a $N_{\rm c}=128^3$ cells regular mesh (corresponding to approximately $l\sim 3.9\,h^{-1}\,\rm{Mpc}$ comoving cell size) and run with a TreePM N-body code scaling as $\mathcal{O}(N_{\rm p}\log(N_{\rm p}))$, \texttt{Hydro-BAM} would keep its performance unchanged whereas the simulation would take approximately $\sim 78$ times the CPU time required by our reference simulation to run up to $z=2$. Lastly, \texttt{BAM} performance does not depend on redshift, whereas it takes a much longer CPU time to evolve the initial conditions to lower redshift, e.g. up to $z\sim1$. In general, there are hence no drawbacks in using larger volume simulations, as long as a reasonable trade-off between number of particles and mesh resolution is guaranteed. In some unfavourable cases, there may not be any further gain in computing time with respect to the reference case investigated in this paper, being anyway left with the 7-orders-of-magnitude computing time gain quoted above, i.e. \texttt{HydroBAM} is not penalized.

When the \HII{} and \HI{} fields are available from the simulation, the bias is modelled with just two conditional properties (\HII{} and \HI{}), so that $N_{\rm bins,tot}\ll N_{\rm cell}$ (or, equivalently, the average number of cells per $\Theta$-bin is $N_{\Theta}=N_{\rm cell}/N_{\rm bins,tot}\gg 1$) and the results are therefore not likely to be affected by overfitting. On the other hand, the case in which \texttt{BAM} learns from reconstructed \HII{} and \HI{} fields, and includes also information on dark matter in the bias model, is more delicate and deserves a more careful look. Because the majority of the bins used to model $\{\Theta\}$ in the bias histogram are actually devoid of cells, following \S4.4 of HydroBAM-I we account only for the bins which are filled with cells instead of considering all the bins used to describe $\Theta$, and define the effective average number of cells per $\Theta$-bin as $N_\Theta^{\rm eff}=N_{\rm cell}/N_{\rm bins,filled}$. The purpose of this is to shift the focus only onto the $\Theta$-bins which are filled with cells and for which is therefore meaningful to define a non-trivial one-dimensional tracer probability distribution, according to which the subsequent random sampling is performed. In other words, considering also the $\Theta$-bins which are devoid of cells would be misleading, as no random assignment of tracers is performed in such cases.\footnote{In the context of \texttt{BAM}, all the $\Theta$-bins used to describe regions of zero probability in non-pathological bias relations are purposeless and can be effectively considered as one single bin. Increasing the binning resolution in such regions helps just in gaining accuracy when defining the transition between zero and non-zero probability regions of the distribution.} This computation yields $N_{\Theta}^{\rm{eff}}\sim 40-50$ depending on the considered iteration within the calibration procedure, which is reasonably high number to regard that we do not incur into overfitting.

From a perturbation theory perspective \citep[see e.g.][]{McDonald2000,McDonald2003,Seljak2012}, this would correspond to a Taylor expansion of the flux field at resolution $\Delta l$, which to linear order is expressed as 
    \begin{eqnarray}
    \lefteqn{F_{\Delta l}/F_{\Delta l}(\delta_{{\rm HII},\Delta l}=\delta_{{\rm HI},\Delta l}=0)=}\\&&1+b_{{\rm HII},\Delta l}\delta_{{\rm HII},\Delta l}({\rm DM})+b_{{\rm HI},\Delta l}\delta_{{\rm HI},\Delta l}({\rm HII}({\rm DM}))\,,\nonumber
    \end{eqnarray}
where convenient expansions for $\delta_{\rm HII}$ and $\delta_{\rm HI}$ can be substituted into the equation to eventually relate it to the dark matter field. In fact, \texttt{BAM} is able to perform linear combinations of the non-linear perturbative Taylor expansions of the properties included in $\lbrace\Theta\rbrace$\citep[see][for a more detailed discussion on the connection between \texttt{BAM} and perturbation theory]{Kitaura2020}.

The novel aspects to the standard perturbative framework are that our bias treatment (i) is  explicitly based on baryon rather than on dark matter densities, (ii) that the dependencies are formulated in a multi-variate fashion, and (iii) that the relations are iteratively learnt from a reference simulation. Separating the \HII{} and \HI{} components, which could be in principle unified into one single baryon density field term, allows to treat them independently, give them different weights, and leave the freedom to build distinct higher order local and non-local terms, to model different baryonic phenomena down to highly non-linear scales.

\section{Reconstruction of the Lyman$-\alpha$ forest: redshift space} \label{sec:rec_redshift}
In this section we show the performance of the \texttt{Hydro-BAM} approach including redshift space distortions.

\subsection{RSD modelling} \label{sec:rsd}

As anticipated in \S\ref{sec:pipeline}, the second part of the calibration pipeline requires the transition from real to redshift space. Given that the \HII{} and \HI{} density fields are reconstructed in real space (as in HydroBAM-I), an explicit modelling of RSD for such quantities is required at this stage. In order to do so, we devise a procedure to apply RSD to density fields already interpolated on a mesh. The procedure we adopt can be summarized in the following steps:
\begin{enumerate}
    \item Assign to each cell $i$ 
    characterized by a local density $\rho_{i}$, a number $N=16$  pseudo-particles\footnote{In general, the performance of our method is found to improve with increasing $N$, though rapidly saturate already at $N=32-64$. The choice $N=16$ represents the optimal trade-off between accuracy and code speed.} with mass $M=\rho_i V_i/N$, and positions $\vec{r}_j$ coincident to the center of the cell; 

    \item Consider the dark matter velocity field $\vec{v}^{\,\rm sim}_{{\rm dm},i}$ from the simulation, interpolated on a mesh of equal size and resolution as the density fields. Apply RSD and transform real-space $\vec{r}_j$ to redshift space $\vec{s}_j$ coordinates by means of the mapping \citep[see e.g.][]{Kaiser1987,Hamilton1998} 
    \begin{equation}\label{r2s}
    \vec{s}_{j}=\vec{r}_{j}+[b_{\rm v}\,(\vec{v}_{{\rm dm},j} \cdot \hat{r}_{j})\,\hat{r}_j]/(aH),
\end{equation}
    where $\vec{v}_{{\rm dm},j}=\vec{v}^{\, \rm coh}_{{\rm dm},j} + \vec{v}^{\, \rm disp}_{{\rm dm},j}$,  with $\vec{v}^{\, \rm coh}_{{\rm dm},j}=\vec{v}^{\,\rm  sim}_{{\rm dm},i}$ 
    is the coherent component of the velocity field, while 
    $\vec{v}^{\, \rm disp}_{{\rm dm},j}$ accounts for the small-scale quasi-virialized motions. This last term is originally smoothed-out from the distribution of gas by the interpolation of particle velocities on the mesh at physical cell-resolution. Therefore, we restore it assuming that its components follow a Gaussian distribution of zero mean and variance $A(1+\delta_i)^\alpha$ with $A$ and $\alpha$ as free parameters \citep{2013MNRAS.435.2065H,2014MNRAS.439L..21K}. The term  $b_{\rm v}$  in Eq.~(\ref{r2s}) is a velocity bias factor, while $a$ is the scale factor, $H$ is the Hubble parameter and $\hat{r}=\vec{r}/|\vec{r}|$. We have adopted the distant observer approximation and translated to redshift space along the z-component.
    \item Interpolate again the redshift-space pseudo-particles on a grid using a CIC MAS, with $M_j$ as weight.

\end{enumerate}
This description of RSD implies a set of three free parameters $\{b_{\rm v},A,\alpha\}$, which are determined by comparing the signal of the multipole decomposition of the $3D$ flux power-spectrum $P_{\ell}(k)$, in particular the quadrupole ($\ell=2$) and hexadecapole ($\ell=4$) \footnote{Multipoles have been measured using \texttt{nbodykit} \citep[\url{https://nbodykit.readthedocs.io},][]{Hand2018}} in Fourier space, leading to $\{b_{\rm v},A,\alpha\} \sim \{0.93,1,1.03\}$. In general, the parameters depend on the grid resolution, and hence these figures might not be the optimal choice in other cases. We will further address this aspect in future works, in which we will apply our method to higher-resolution mesh representations of the same reference simulation.

\begin{figure*}
\centering
\includegraphics[width=18cm]{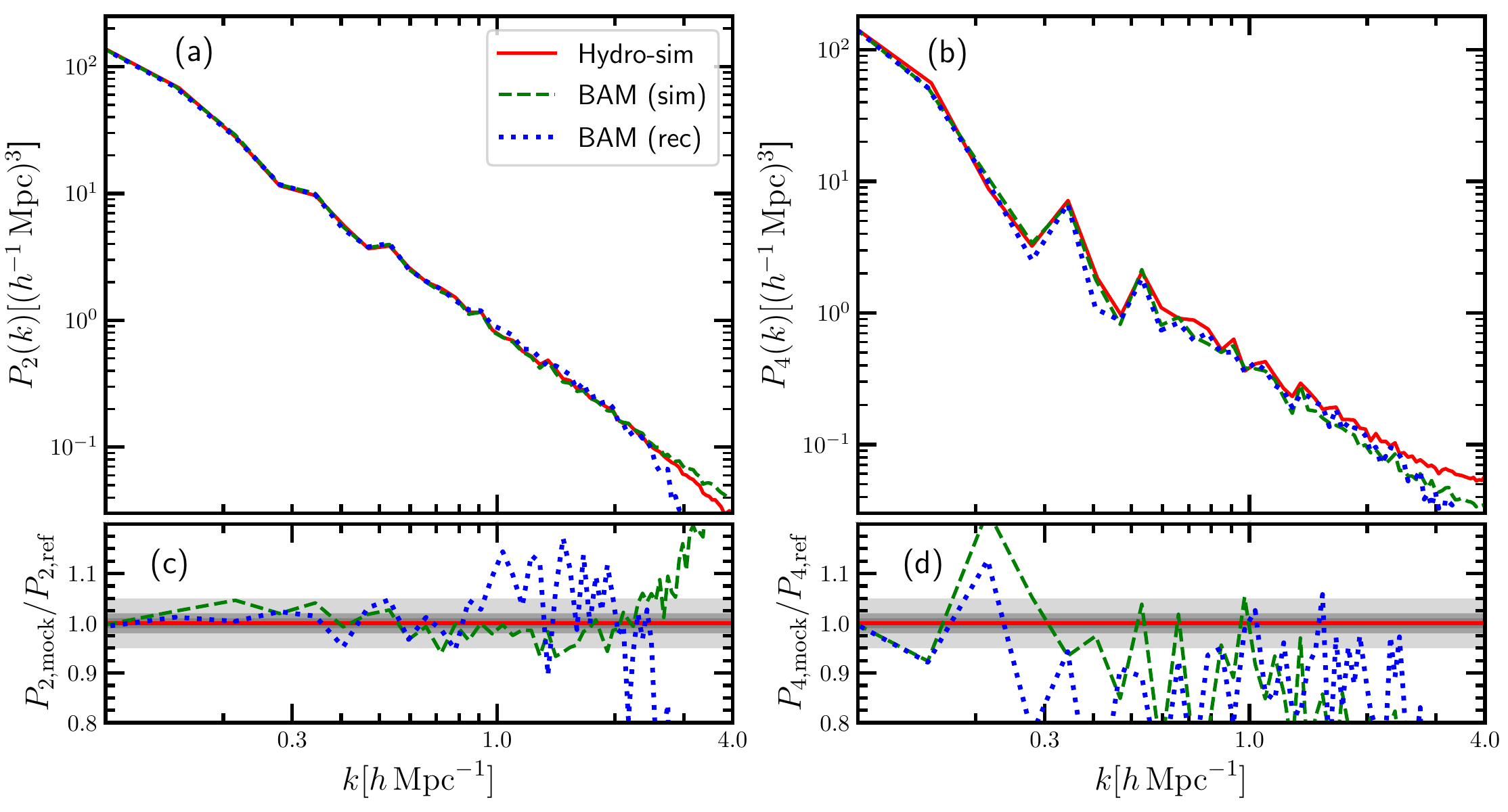}
\caption{\small{Multipole decomposition of the \Lyaf{} in Fourier space. Panels (a) and (b) show respectively the quadrupole $P_{2}(k)$ and the hexadecapole $P_{4}(k)$ measured from the reference simulation, \texttt{BAM} reconstruction using density fields from the reference simulation and the \texttt{BAM} reconstruction using \texttt{BAM}-reconstructed density fields. The bottom panels (c) and (d) show the ratios between the multipoles form the two \texttt{BAM} reconstructions and the reference simulation. In these panels, gray shaded area in indicate the $1\%,2\%$ and $5\%$ deviations from the reference simulation.}}
\label{fig:multipoles}
\end{figure*}

\subsection{Reconstruction} 

The final phase of our pipeline consists of reconstructing the \Lyaf{} field in redshift space, with the bias characterized by the set of underlying properties  $\{\Theta\}=\{\delta_{\rm{HII}}(\vec{s})\otimes\mathcal{K},\delta_{\rm{HI}}(\vec{s})\}$.

We first apply \texttt{BAM} to the reference \HII{} and \HI{} density fields (mapped in redshift space through $\mathcal{S}$), to address the performance of our RSD model in the bias. Subsequently, we repeat the same using \texttt{BAM}-reconstructed \HII{} and \HI{} fields, to mimic a realistic mock production context.

Similarly to the case of real space, \texttt{BAM} is found to yield accurate results. Dedicated experiments reveal that in the latter the outcome suffers larger deviations than expected. This is due to the progressive loss of accuracy at $k>1.0\,h\,\rm{Mpc}^{-1}$ affecting the reconstructed \HII{} and \HI{} fields. Therefore, we re-introduce the dependence in the bias on the dark matter field ($\{\Theta'(\vec{s})\}=\{\delta^z_{\rm{HII}}(\vec{s})\otimes\mathcal{K},\delta_{\rm{HI}}(\vec{s}),\delta_{\rm{dm}}(\vec{s})\}$, as previously discussed in\S\ref{sec:pipeline}), conveniently mapped in redshift space as well.


The preliminary qualitative assessment based on the visual inspection of slices through the simulation box (Fig. \ref{fig:slices}, bottom row) highlights  that, the same as in real space, \texttt{BAM} (central panel) succeeds in reproducing the small- and large-scale structure of the reference slice (left panel) with a high level of agreement.

A more quantitative assessment can be performed through the analysis of the reconstructed summary statistics. Panel (d) in Fig.~\ref{fig:pdfs} shows that both the \Lyaf{} field reconstructions built on the simulation-based (blue dashed) and \texttt{BAM}-based (brown dash-dotted) \HII{} and \HI{} fields, are found to have extremely accurate PDF if compared to the reference (red solid). 

Panel (a) in Fig.~\ref{fig:power_spectra} shows the power-spectrum reconstruction of the flux field performed by \texttt{BAM}, based on \HII{} and \HI{} density fields from the simulation (blue dashed) and from previous \texttt{BAM} runs (brown dash-dotted) respectively, compared to the reference flux field power-spectrum (red solid). The panel (b) of the same figure shows the ratio $P_{\rm mock}(k)/P_{\rm ref}(k)$, indicating that in this setup \texttt{BAM} is able to learn the bias relation from the simulation reproducing the reference power-spectrum with average deviations $\lesssim 1.7\%$ ($<1\%$) and maximum deviation $\sim3\%$ ($\sim 2\%$) up to the Nyquist frequency (up to $k=1.0\,h\,\rm{Mpc}^{-1}$). Learning the bias relation from baryon fields previously reconstructed by \texttt{BAM} yields a moderate loss of accuracy at $k>1.0\,h\,\rm{Mpc}^{-1}$, albeit preserving the accuracy on large-scales, accounting for average and maximum deviations $\lesssim 2\%$ and $\sim3.5\%$ respectively at $k\lesssim1.0\,h\,\rm{Mpc}^{-1}$. As mentioned, the average mild lack of power trend at $k\gtrsim1.0\,h\,\rm{Mpc}^{-1}$, dominated by superimposed random noise oscillations, is due to the progressive loss of accuracy in \HII{} and \HI{} reconstructions, which unavoidably propagates to the \Lyaf{} as well. In fact, modelling the power spectrum at $k > 1.0\, h\, \rm{Mpc}^{-1}$ (where AGN and supernova feedback have a significant effect) in redshift space is a highly non-trivial task. Hence, to some extent it is not surprising that a loss of accuracy in \texttt{BAM} reconstructions of \HII{} and \HI{} has a negative impact also on the subsequent reconstruction of the \Lyaf{}, based on the previous two (see also \S\ref{sec:bam_lim}).

To further assess the accuracy of our reconstruction and the performance of RSD modelling, we compute the multipole expansions of the power-spectrum. In particular,  the quadrupole and the hexadecapole of \Lyaf{} reconstructions are shown in panels (a) and (b) of Fig.~\ref{fig:multipoles},  respectively. While the power-spectrum is constrained in \texttt{BAM} through the iterative procedure, the quadrupole and hexadecapole are not explicitly constrained. Therefore, the accuracy in the reproduction of such multipoles is primarily a consequence of the approach we have adopted for RSD. The reference quadrupole is shown to be recovered with $<5\%$ maximum deviation at $k\lesssim 1.0\,h\,\rm{Mpc}^{-1}$, for reconstructions based both on the reference simulation (green dashed) and on \texttt{BAM} baryon density fields (purple dash-dotted). Average deviations of the hexadecapole with respect to the reference (red solid) are $\sim10\%$ and $\sim 15\%$ at $k\lesssim 0.5\,h\,\rm{Mpc}^{-1}$ when learning from the simulation gas (blue dashed) and from the \texttt{BAM} gas (brown dash-dotted), respectively.

The bottom plots set in Fig. \ref{fig:bi-spectra} shows the reconstructed flux bi-spectrum at different triangular configurations in Fourier space. \texttt{BAM} credibly reproduces the reference bi-spectrum (red solid) at $k\lesssim 0.8\,h\,\rm{Mpc}^{-1}$ when building on the \HII{} and \HI{} fields both from the simulation (blue dashed) and from reconstructions (brown dash-dotted), starting to lose accuracy in $\theta_{12}\approx[0,1.3]$ only at $k>0.8 \, h\,\rm{Mpc}^{-1}$.

As in the case of real space, an assessment of overfitting is needed only in the case of reconstruction based on \HII{} and \HI{} fields obtained by \texttt{BAM}, in which the dark matter field is included as well in the bias. Again, $N_{\Theta}^{\rm{eff}}\sim40-50$ and we can rule out the possibility that overfitting is affecting our results.  

We consider the RSD modelling acceptable at this stage. A number of modifications can be introduced to improve the accuracy in the  multipoles at the expense of either introducing more parameters, or another machine learning step.

For the first strategy, we could consider separate $b_{\rm v}$ velocity bias for \HII{}, \HI{}, and the DM. Also the corresponding dispersion terms could adopt  different $A$ and $\alpha$ values. These improvements may allow to account for intrinsic velocity bias differences among these fields. We do not expect these refinements to introduce dramatic changes in the RSD free parameters we have estimated in this work. However, leaving the algorithm the freedom of choosing different values for the parameters related to different fields might improve even further the RSD modelling, with a consequent potential projected improvement in the accuracy of our reconstructions.

In a second strategy, we could introduce \texttt{BAM}-like additional machine learning steps to extract the degree of anisotropy as a function of scale and density required  to minimize a cost function determined by, e.g., the quadrupole or hexadecapole. In particular, the parameters of our RSD model would be automatically determined in a machine learning fashion by explicitly requiring that deviations of higher order multipoles from the reference statistics are minimized in the cost function, currently not implemented in our algorithm. Concretely, in this case the new machine learning step would consist in adding a further optimization stage in the \texttt{BAM} flowchart sketched in \S\ref{sec:pipeline}, repeating iteratively steps 3, 4, and 5, to determine the values of $b_v$, $A$ and $\alpha$ which optimize the quadrupole and hexadecapole. Extracting such parameters as a function of scale and density might allow to unveil non-obvious dependencies and, thus, improve the model.

We leave these series of studies for later work.
\newpage
\section{Limitations of the method} \label{sec:bam_lim}

In what follows, we summarize the main potential drawbacks and limitations of our method. 
\begin{itemize}
    \item Availability of a suitable reference simulation: since \texttt{BAM} is in principle able to self-consistently learn the kernel from just one, or few, reference simulations, the main limitation to its application is the availability of such a suitable simulation. In machine learning language, such limitation translates into the availability of a suitable, sufficiently large, training dataset. While this condition is unavoidable by construction and represents a cause of likely failure if not fulfilled, as already mentioned \texttt{Hydro-BAM} alleviates the need for a rather large training set (i.e. tens of simulations), commonly shared by other machine learning algorithms through an explicit modelling of the physical dependencies underlying the bias relation. In this sense, as already mentioned in \S\ref{sec:bam_ml}, one or few simulations spanning a sufficiently large total volume, are enough to guarantee that the learning processes is successful and not prone to cosmic variance effects.  A volume in our context is sufficiently large when: \begin{enumerate}
        \item the mode coupling on large scales is correctly included \citep[][]{Crocce2006,Kyplin2018}, which agrees with a flat kernel towards low $k$ values \citep[][]{Balaguera2019};
        \item the number of cells exceeds the number of parameters of the bias model (i.e., the total number of bins) to ensure good enough statistics for each bias dependency.
    \end{enumerate} 
     \item Need for an effective and feasible prescription for the bias: while the architecture of \texttt{Hydro-BAM} --- relying on an explicit modelling of the key physical dependencies in the bias relation --- has been proven to be a major strength of our method, it can become a drawback if one has little clue of which dependencies are the leading ones.   This is particularly true for non-local bias dependencies. While a common ML method is able to explore any kind of non-local dependence without understanding the underlying physics, the \texttt{Hydro-BAM} approach needs some explicit information such as the tidal field tensor. An extension of this problem is that \texttt{Hydro-BAM} needs in principle all relevant physical relations given to reach a certain accuracy.  This drawback could become especially problematic when going to smaller scales, when we need to specify not only the tidal field tensor, but also the density fluctuation curvature, or even the velocity shear. A ML algorithm can potentially automatically explore all dependencies and even blindly handle them with multiple layers. For the considered scales in this work, we consider nonetheless that \texttt{Hydro-BAM} is very competitive and could inspire ML algorithms. 
     Even though all the applications shown thus far in \texttt{BAM} papers are supported by a long-standing theoretical discussion in the literature, offering valuable insights into the topic, this might not be the case for other applications not yet explored. Moreover, a similar issue arises if the bias relation requires the modelling of quantities beyond the dark matter density field, which cannot be reliably obtained within \texttt{BAM}. In this case, one can still extract the needed fields from the reference simulations to carry out a purely academic study, but will not be able to exploit such information to generate mock catalogs. E.g., in this work, we make use of the \texttt{BAM}-reconstructed \HII{} and \HI{} fields to map the \Lyaf{}. If we had found other gas properties to play a crucial role in the \Lyaf{} mapping, we would have had to go beyond the hierarchical approach presented in HydroBAM-I and devise new strategies. 
    \item Loss of accuracy in the various \texttt{BAM} hierarchical reconstruction steps: we have shown in this paper that the application of the whole \texttt{Hydro-BAM} pipeline results in a very accurate reconstruction in redshift space of the power spectrum up to $k\sim 1.0\,h^{-1}\,\rm{Mpc}$, and of the bi-spectrum up to $k\sim 0.8\,h^{-1}\,\rm{Mpc}$. The loss of accuracy at $k\gtrsim 1.0\,h^{-1}\,\rm{Mpc}$ is a natural consequence of the fact that  baryon processes have an important impact in determining the \Lyaf{} power spectrum at such frequencies. Therefore, achieving high accuracy in such a highly-nonlinear regime is very non-trivial, and a loss of accuracy in the various reconstruction steps progressively forward-propagate until the last step.  
    \item Dependence on cosmological and astrophysical models of the reference simulation: as \texttt{BAM} extracts the relations linking different quantities from the reference simulations, where such quantities have been obtained through the full $N$-body/SPH calculation, the resulting \texttt{BAM} mock catalogs will forcefully represent cosmic realizations bound to the input parameters and models adopted to run the reference training simulation. This implies that, to perform predictions with different cosmological models (e.g. modified gravity), or different baryon feedback implementations, it will be necessary to dispose of a suite of different training simulations, one for each different sets of models and parameters to be investigated.
\end{itemize}

\newpage
\section{Conclusions} \label{sec:conclusions}

In this paper, we have presented a hierarchical domain specific machine learning approach to efficiently map the three-dimensional  \Lyaf{} field starting from the dark matter density field.
We have shown that a direct relation cannot be accurate if we ignore the \HII{} and \HI{} distributions.
Therefore, the machine learning approach has to be first applied to obtain the \HII{} field, which shows the highest correlation with the dark matter field, and then in turn that baryonic field has to be used to get the \HI{} field, as described in \cite{Sinigaglia2020}.
These fields are  combined in a non-linear,  non-local multivariate fashion to obtain the \Lyaf{}  absorption flux field. In this way, a few percent accuracy is achieved in the summary statistics.

Our results, first obtained in real space, show maximum and average deviations in the power-spectrum up to the Nyquist frequency of $\sim 3\%$ and $<1.7\%$, respectively; and good agreement in the bi-spectra  probing scales down to $k\sim1.0\,h\,\rm{Mpc}^{-1}$. In contrast, the commonly-adopted FGPA reproduces the \Lyaf{} power-spectrum with  oscillations of $\sim5\%$ around the reference up to $k\sim 1.0\,h\,\rm{Mpc}^{-1}$, failing to match the one-point PDF and is found to progressively lose accuracy in the bi-spectrum at configurations already with $k>0.1\,h\,\rm{Mpc}^{-1}$.

In order to represent the \Lyaf{} in redshift space (as probed by galaxy surveys), we map the \HII{} and \HI{} fields from real to redshift space through an appropriate modelling of RSD, not included in HydroBAM-I. In particular, we explicitly account for both large-scale coherent flows and small-scale quasi-virialized motions contributions. 

Having these consistent anisotropic fields we apply the same  machine learning scheme as in real space to reconstruct the \Lyaf{} field in redshift space. 

The reconstructed \Lyaf{} power-spectrum presents average deviations of $\lesssim 1.7\%$ ($\lesssim2\%$) and maximum deviations of $\sim 2\%$ ($\sim 3\%$) up to $k\sim 1.0\,h\,\rm{Mpc}^{-1}$ (up to the Nyquist frequency) when the reconstruction of fluxes is based on \HII{} and \HI{} density fields previously reconstructed with \texttt{BAM} following HydroBAM-I (from the simulation). The reconstructed quadrupole deviates $<5\%$ with respect to that from the reference at $k<0.9\,h\,\rm{Mpc}^{-1}$ when using \texttt{BAM}-reconstructed baryon densities and up to $k\sim2.0\,h\,\rm{Mpc}^{-1}$ when relying on baryon densities from the reference simulation. The hexadecapole reproduces the reference with reasonable accuracy considering that it is highly affected by cosmic variance. Finally, the bi-spectrum is well-reproduced for Fourier triangular configurations with side lenghts up to $k\sim 0.8\,h\,\rm{Mpc}^{-1}$.

We stress that our hierarchical bias mapping is applied for the first time in this paper also using in the calibration \HII{} and \HI{} density fields self-consistently reconstructed within \texttt{BAM}, whereas in HydroBAM-I we showcased only the ideal situation in which such fields can be read from the simulation.

In forthcoming publications we will extend the technique presented in this work to a larger volume simulation, aiming at producing mock \Lyaf{} realizations to cover the cosmic volumes probed by on-going and future observational campaigns. In parallel, we will address the bias modelling towards smaller scales, to improve the accuracy of reconstructed quantities in \texttt{Hydro-BAM} on scales where relevant astrophysical phenomena, such as feedback and cooling, plays a major role in determining the spatial distribution of baryons. As mentioned, in this context we also plan to extend our study to higher-resolution mesh representation of the same reference simulations, to get \texttt{Hydro-BAM} ready to explore new avenues and study phenomena which can impact the \Lyaf{} summary statistics on very small scales ($k\sim5-10\,h\,\rm{Mpc}^{-1}$), such as the mass of warm dark matter particles \citep{Viel2005,Viel2013a}, the mass of neutrinos \citep{PalanqueDelabrouille2015} and the thermal state of the intergalactic medium \citep[][and references therein]{Garzilli2017}, among others.

Furthermore, our method can be applied in a similar fashion as in this paper to obtain independent \HI{} density fields in redshift space which, if properly complemented with instrumental effects, can pave the way to the generation of accurate and detailed 21cm-line mock catalogs for \HI{} intensity mapping. 

We stress that \texttt{BAM} as a particularly efficient machine learning technique has still plenty of potential applications to be explored in the fields of large-scale structure and galaxy formation and evolution, limited a priori just by the availability of a suitable reference simulation.

\section*{Acknowledgments}

We acknowledge the anonymous referee for their comments, which have helped to improve the quality of the manuscript. We warmly thank Marc Huertas-Company for valuable comments and insights on machine learning.
FS acknowledges the support of the doctoral grant funded by the University of Padova and by the Italian Ministry of Education, University and Research (MIUR). FS is also grateful to the \emph{Instituto de Astrof\'{\i}sica de Canarias} (IAC) for hospitality and computing resources.
FSK and ABA acknowledge the IAC facilities and  the Spanish Ministry of Economy and Competitiveness (MINECO) under the Severo Ochoa program SEV-2015-0548, AYA2017-89891-P and CEX2019-000920-S grants. FSK also thanks the  RYC2015-18693 grant.
K.N. is grateful to Volker Springel for providing the original version of {\sc GADGET-3}, on which the {\sc GADGET3-Osaka} code is based on. Our numerical simulations and analyses were carried out on the XC50 systems at the Center for Computational Astrophysics (CfCA) of the National Astronomical Observatory of Japan (NAOJ), the XC40 and the Yukawa-21 at YITP in Kyoto University, {\sc Octopus} at the Cybermedia Center, Osaka University, and {\small Oakforest-PACS} at the University of Tokyo as part of the HPCI system Research Project (hp200041, hp210090).  This work is supported in part by the JSPS KAKENHI Grant Number JP17H01111, 19H05810, 20H00180.
K.N. acknowledges the travel support from the Kavli IPMU, World Premier Research Center Initiative (WPI), where part of this work was conducted. 
MA thanks for the support of the Kavli IPMU fellowship and was supported by JSPS KAKENHI Grant Number JP21K13911. We acknowledge Cheng Zhao for making available his code to measure the bi-spectrum. 

\vspace{1cm}

\bibliography{sample}{}
\bibliographystyle{aasjournal}

\end{document}